\documentclass[twocolumn,prb,floatfix,superscriptaddress]{revtex4-1}
\usepackage{amsfonts}
\usepackage{amsmath,amsbsy,amssymb}
\usepackage[dvipdfmx]{graphics}
\usepackage{float}
\usepackage{bm}
\usepackage{braket}
\usepackage{ulem}
\usepackage{mathtools}
\usepackage[breaklinks,colorlinks=true,linkcolor=blue,urlcolor=blue,citecolor=blue]{hyperref}
\usepackage{verbatim}

\begin{document}

\title{Higher-order topological phases in a spring-mass model on a breathing kagome lattice}
\author{Hiromasa Wakao}
\affiliation{Graduate School of Pure and Applied Sciences, University of Tsukuba, Tsukuba, Ibaraki 305-8571, Japan}
\email{wakao@rhodia.ph.tsukuba.ac.jp}
\author{Tsuneya Yoshida}
\affiliation{Department of Physics, University of Tsukuba, Tsukuba, Ibaraki 305-8571, Japan}
%\email{yoshida@rhodia.ph.tsukuba.ac.jp}
\author{Hiromu Araki}
\affiliation{Graduate School of Pure and Applied Sciences, University of Tsukuba, Tsukuba, Ibaraki 305-8571, Japan}
%\email{araki@rhodia.ph.tsukuba.ac.jp}
\author{ Tomonari Mizoguchi}
\affiliation{Department of Physics, University of Tsukuba, Tsukuba, Ibaraki 305-8571, Japan}
%\email{mizoguchi@rhodia.ph.tsukuba.ac.jp}
\author{Yasuhiro Hatsugai}%
%\email{hatsugai@rhodia.ph.tsukuba.ac.jp}
\affiliation{Department of Physics, University of Tsukuba, Tsukuba, Ibaraki 305-8571, Japan}

\date{\today}
\begin{abstract}
We propose a realization of higher-order topological phases in a spring-mass model with a breathing kagome structure.
To demonstrate the existence of the higher-order topological phases, we characterize the topological properties and show that the corner states appear under the fixed boundary condition.
To characterize the topological properties, we introduce a formula for the $\mathbb{Z}_3$ Berry phases in the Brillouin zone.
From the numerical result of this $\mathbb{Z}_3$ Berry phase, we have elucidated that coupling between the longitudinal and transverse modes yields a state characterized by the Berry phase $\frac{2\pi}{3}$ for our mechanical breathing kagome model.
In addition, we suggest that the corner states can be detected experimentally through a forced vibration.
\end{abstract}

\maketitle

%\input{introduction.tex}
%Introduction
\section{Introduction}

%Introduction of conventional TIs
Topological insulators (TIs)~\cite{RevModPhys.82.3045,RevModPhys.83.1057}
are distinctive class of insulators where topologically-nontrivial structures of Bloch wave functions
give rise to characteristic boundary states.
Various unique phenomena are caused by these boundary states,
such as
quantization of Hall~\cite{PhysRevLett.61.2015} and spin-Hall conductivities~\cite{PhysRevLett.95.226801},
and electromagnetic responses~\cite{PhysRevB.78.195424,PhysRevLett.102.146805}.
In TIs, bulk topological invariants which characterize the nontrivial topology of Bloch wave functions
are known to be related to the $(d-1)$-dimensional boundary states
(with $d$ being spatial dimension of the bulk Hamiltonian); this relation is nowadays established as bulk-boundary correspondence~\cite{PhysRevLett.71.3697,PhysRevB.48.11851}.

%What are HOTIs
Recently, a novel class of TIs, called higher-order topological insulators (HOTIs), was introduced~\cite{Hayashi2018,Benalcazar61,PhysRevB.96.245115,PhysRevLett.120.026801,Schindlereaat0346,PhysRevB.97.241405,2017arXiv171109202X,PhysRevB.99.041301}.
In HOTIs, $d-2$ or fewer dimensional boundary states appear in $d$-dimensional models,
which is predicted by topological invariants in the bulk.
Examples of such topological invariants include the multipole moment~\cite{Benalcazar61,PhysRevB.100.245134}, the nested Wilson loops~\cite{PhysRevB.96.245115,2017arXiv171109202X},
the quantized Wannier centers~\cite{PhysRevLett.120.026801},
the entanglement polarization~\cite{PhysRevB.98.035147},
and the $\mathbb{Z}_Q$ Berry phase~\cite{Hatsugai_2011,PhysRevLett.123.196402,PhysRevResearch.2.012009}.
In that sense, a novel kind of bulk-boundary correspondence emerges in HOTIs.

% The mechanical models realizing HOTIs
In parallel with the theoretical developments, realization of HOTIs in solids has actively been pursued~\cite{PhysRevB.98.045125,schindler2018higher,PhysRevLett123256402,2019arXiv190411452L,2019arXiv190700012L,PhysRevMaterials.3.114201}.
In addition, higher-order topological phases in artificial systems have also been studied intensively.
These systems are advantageous compared with solids from the viewpoints of
simplicity of experimental set-up and high tunability of parameters,
which enable us to implement desirable structures to realize higher-order topological phases.
Indeed, the higher-order topological phases were realized in mechanical systems~\cite{serra2018observation,PhysRevResearch.1.032047}, photonic crystals~\cite{ElHassan2019,Ota:19}, phononic crystals~\cite{Ni_2017,xue2019acoustic,ni2019observation}, electrical circuits~\cite{imhof2018topolectrical,PhysRevB.98.201402}, and carbon monoxide molecules on a Cu(111) surface~\cite{Kempkes2019observation}.

%What is topological phenomena in this mechanical model? &&
In this paper, we propose a realization of higher-order topological phases in a spring-mass model.
Spring-mass models, composed of a periodic alignment of springs and mass points,
serve as a simple platform to realize topological phenomena governed by Newton's equation of motion ~\cite{kane2014topological,Wang_2015,kariyado2015manipulation,takahashi2017edge,PhysRevB.99.024102,PhysRevB.100.054109}.
Indeed, topological phases accompanied by characteristic boundary states, such as Chern insulators~\cite{Wang_2015,kariyado2015manipulation}, nodal-line semimetals~\cite{takahashi2017edge}, and Weyl semimetals~\cite{PhysRevB.99.024102}, were proposed.
These results motivated us to study the higher-order topological phase in a spring-mass model.

%specifically what I did
%What I study
As a concrete example, we study the spring-mass model on a breathing kagome lattice, and demonstrate the realization of the higher-order topological phase in our model.
Specifically, we characterize the topological phases by introducing a formula for the $\mathbb{Z}_3$ Berry phase in the Brillouin zone and the existence of the corner states.
In addition, from this numerical result of this bulk properties, we have elucidated that coupling between the longitudinal and the transverse modes yields a state characterized bt the Berry phase $\frac{2\pi}{3}$ for our mechanical breathing kagome model.
We further propose how to observe the corner states experimentally.
To this aim, we study the dynamics under the external force and show the characteristic behavior of corner states distinct from the bulk states.

%What we show in this paper
The rest of this paper is organized as follows.
In Sec.~\ref{two}, we introduce the spring-mass model on a breathing kagome lattice and explain how to describe the motion of mass points in this model.
In Sec.~\ref{three}, we first explain the definition of the $\mathbb{Z}_{3}$
Berry phases in momentum space.
We then show the numerical results for the bulk properties such as band structures and $\mathbb{Z}_{3}$ Berry phases.
In Sec.~\ref{four}, we elucidate the existence of the corner states in this model under the fixed boundary conditions with a triangle arrangement.
In Sec.~\ref{five}, we demonstrate that the corner states can be observed by the forced vibration.
In Sec.~\ref{six}, we present a summary of this paper.
In Appendix~\ref{eight}, we see the $\mathbb{Z}_{3}$ Berry phase for the lower five bands, which accounts for the corner states under the weak tension.
In Appendix~\ref{nine}, we show the band structure on the cylinder geometry.
In Appendix~\ref{ten}, we show the numerical result of the inverse participation ratio (IPR) to see the corner states from the bulk or edge continuum.
In Appendix~\ref{eleven}, we show the existence of the corner state under the fixed boundary conditions with in a parallelogram arrangement.

%\input{spring-mass.tex}
%Introduction the spring-mass model on a breathing kagome
\section{a spring-mass model on a breathing kagome lattice}
\label{two}
%------SET UP
%%%%%%%%%%%%
\begin{figure}[htb]
  \includegraphics[width=\linewidth]{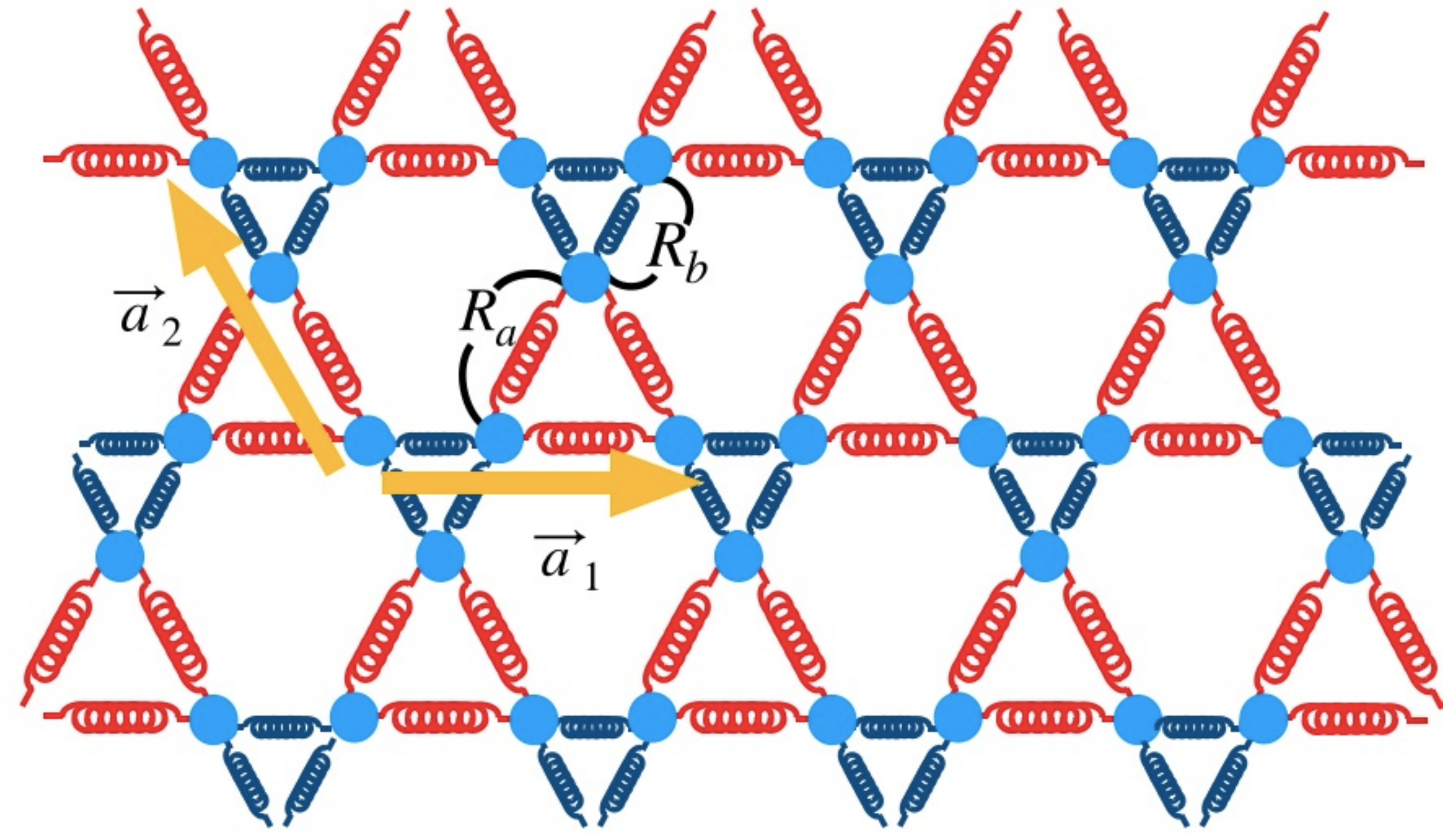}
\caption{
A spring-mass model in a breathing kagome structure.
}
 \label{spring-mass_model}
\end{figure}
%%%%%%%%%%%%

We consider a system consists of the mass points aligned on a breathing kagome lattice and springs connecting the masses (Fig.~\ref{spring-mass_model}).
The spring constants of red springs on upward triangles are $t_{a}$, and those of blue springs on downward triangles are $t_{b}$.
We label a red spring (a blue spring) as $\alpha=a$ ($b$).
Henceforth, we set the mass as unity for simplicity.

We set unit vectors $\vec{a}_{1}$ and $\vec{a}_{2}$ as
%%%%%%%%%%%%
\begin{subequations}
%%%
\begin{equation}
\vec{a}_{1}=
(R_{a}+R_{b})
\left(
 \begin{array}{c}
  1\\
  0
 \end{array}
\right)\label{unit1}
\end{equation}
%%%
and
%%%
\begin{equation}
 \vec{a}_{2}
=(R_{a}+R_{b})
\left(
\begin{array}{c}
 -\frac{1}{2}\\
\frac{\sqrt{3}}{2}
\end{array}
\right),
\label{unit2}
\end{equation}
%%%
\end{subequations}
%%%%%%%%%%%%
where $R_{a}$ and $R_{b}$ denote lengths of red springs and blue springs in equilibrium, respectively.
To make the model be in equilibrium, we have to take into account a balances of forces, which is satisfied for
%%%%%%%%%%%%
\begin{equation}
 t_{a}(R_{a}-l_{a})=t_{b}(R_{b}-l_{b}),
\label{equiv}
\end{equation}
%%%%%%%%%%%%
where $l_{\alpha}$ is the natural length of a spring.
In the following, we set $l_a = l_b = 1$.

%----- How to describe the motion of spring-mass models
Let us see how to describe the motion of mass points.
Dynamical variables are  $\vec{x}_{\vec{R},p}$,
which are displacements of the mass points from the position in equilibrium.
Here, a pair of indices $\vec{R}$ and $p$ specifies the lattice points;
$\vec{R}$ denotes the position of the unit cell,
and $p=1,2,3$ denotes a sublattice.
For later use, we introduce $\vec{r}_{p}$, which denotes the position of sublattice $p$ within the unit cell.
The Lagrangian describing the motion of masses is written as~\cite{kariyado2015manipulation}
\begin{equation}
\begin{split}
 \mathcal{L}&=\frac{1}{2} \sum_{\vec{R}}\sum_{p} \left(\dot{x}^{\mu}_{\vec{R},p}\right)^{2} \\
  & \quad -\frac{1}{2} \sum_{\langle\vec{R},p;\vec{R}',q\rangle} (x^{\mu}_{\vec{R},p}-x^{\mu}_{\vec{R}',q})\gamma_{\vec{R}+\vec{r}_{p}-\vec{R}'-\vec{r}_{q}}^{\mu\nu} (x^{\nu}_{\vec{R},p}-x^{\nu}_{\vec{R}',q}),
\end{split}
\label{Lagrangian}
\end{equation}
where $\langle\vec{R},p;\vec{R}',q\rangle$ means nearest-neighbor pairs of the mass points, $\mu,\nu=x,y$
are directions in a two dimensional space,
and the explicit form of $\gamma^{\mu\nu}_{\vec{R}+\vec{r}_{p}-\vec{R}'-\vec{r}_{q}}$ is $\gamma^{\mu\nu}_{\vec{R}+\vec{r}_{p}-\vec{R}'-\vec{r}_{q}}=t_{\alpha}\left\{(1-\eta_{\alpha})\delta_{\mu\nu}+\eta_{\alpha} \hat{X}^{\mu} \hat{X}^{\nu}\right\}$.
Here, $\vec{X}=\vec{R}+\vec{r}_{p}-\vec{R}'-\vec{r}_{q}$, and $\hat{X}^{\mu}=X^{\mu}/|\vec{X}|$. The parameter $\eta_{\alpha}$ is defined as
\begin{equation}
 \eta_{\alpha}=\frac{l_{\alpha}}{R_{\alpha}},
\end{equation}
which denotes the strength of tensions of springs.
Notice that the index $\alpha$ in $\gamma^{\mu\nu}_{\vec{R}+\vec{r}_{p}-\vec{R}'-\vec{r}_{q}}$
is naturally determined once we specify the nearest-neighbor pair $\langle \vec{R},p; \vec{R}',q \rangle$.
The first term in Eq.~(\ref{Lagrangian}) is the kinetic energy, while the second term is the potential energy of the springs.

Then, the Euler-Lagrange equation for $x^{\mu}_{\vec{R},p}$,
\begin{equation}
\frac{d}{dt} \left(\frac{\partial \mathcal{L}}{\partial \dot{x}^{\mu}_{\vec{R},p} } \right)- \frac{\partial \mathcal{L}}{\partial x^{\mu}_{\vec{R},p} }=0,
\end{equation}
can be written as a coupled differential equation,
\begin{equation}
 \label{real}
\ddot{\vec{x}}+D \vec{x}=0,
\end{equation}
where $\vec{x}$ is the column vector obtained by aligned $x^{\mu}_{\vec{R},p}$. The matrix $D$,
\begin{equation}
\label{Dynamical_matrix}
\begin{split}
 (D)_{\vec{R},p,\mu;\vec{R}',q,\nu}&=\left(\sum_{\langle \vec{R},p;\vec{R}''o \rangle}\gamma^{\mu\nu}_{\vec{R}+\vec{r}_{p}-\vec{R}''-\vec{r}_{o}} \right)\delta_{\vec{R},\vec{R}'}\delta_{p,q} \\
&\hspace{2.8cm} -\gamma^{\mu\nu}_{\vec{R}+\vec{r}_{p}-\vec{R}'-\vec{r}_{q}},
\end{split}
\end{equation}
is a real-space dynamical matrix.
Assuming the mass points oscillate with a frequency $\omega$,
we can write $x^{\mu}_{\vec{R},p}= e^{i\omega t}\xi^{\mu}_{\vec{R},p}$. Substituting this into Eq.~(\ref{real}), we obtain
\begin{equation}
\label{eigenreal}
-\omega^{2}\vec{\xi}+D\vec{\xi}=0.
\end{equation}
Equation~(\ref{eigenreal}) is an eigenvalue equation of the matrix $D$ whose basis is $\xi_{\vec{R},p}^{\mu}$ and eigenvalue is $\omega^{2}$.
This equation describes the motion of masses in a spring-mass model in real space.

Under the periodic boundary condition, the translational invariance of the system results in the eigenvalue equation in the momentum space. First, we apply the Fourier transformation
%%%%%%%%%%%%
\begin{equation}
\label{Fourier}
 x^{\mu}_{\vec{R},p}=\frac{1}{N}\sum_{\vec{k}} e^{i\vec{k}\cdot \vec{R}} u^{\mu}_{\vec{k},p}.
\end{equation}
%%%%%%%%%%%%
Substituting Eq.~(\ref{Fourier}) into Eq.~(\ref{Lagrangian}), the Lagrangian is written as
%%%%%%%%%%%%
\begin{subequations}
\begin{equation}
\mathcal{L}=\frac{1}{N}\sum_{\vec{k}} \left\{\frac{1}{2} \sum_{p} \dot{u}^{\mu}_{\vec{k},p}\dot{u}^{\mu}_{-\vec{k},p} -\frac{1}{2} \sum_{pq} \Gamma^{\mu\nu}_{pq}(\vec{k}) u^{\mu}_{\vec{k},p} u^{\nu}_{-\vec{k},q} \right\},
\label{bulk_Lagrange}
\end{equation}
%%%%%%%%%%%%
with
\begin{equation}
\label{bulk-dynamical_matrix}
\Gamma^{\mu\nu}_{pq} (\vec{k}) =\sum_{\vec{R}} (D)_{\vec{0},p,\mu;\vec{R},q,\nu}e^{i\vec{k}\cdot \vec{R}}.
\end{equation}
\end{subequations}
The matrix $\Gamma$ is called a momentum-space dynamical matrix.
The dimension of the momentum-space dynamical matrix in this model is six since there are three sublattices and two spatial coordinates.
Under the basis $\left(u^{x}_{\vec{k},1},u^{y}_{\vec{k},1},u^{x}_{\vec{k},2},u^{y}_{\vec{k},2},u^{x}_{\vec{k},3},u^{y}_{\vec{k},3}\right)^{\rm T}$,
the explicit form of $\Gamma(\vec{k})$ is
%%%%%%%%%%%%
\begin{subequations}
\begin{equation}
\begin{split}
 &\Gamma=
  \begin{pmatrix}
   D_{1} & -\gamma_{12}(\vec{k}) & -\gamma_{13}(\vec{k})\\
   -\gamma_{12}^{\dagger}(\vec{k}) & D_{2} & -\gamma_{23}(\vec{k}) \\
   -\gamma_{13}^{\dagger}(\vec{k}) & -\gamma_{23}^{\dagger}(\vec{k}) &D_{3}
  \end{pmatrix}
\end{split},
\end{equation}
%
%
%
%txtb{
%
with
\begin{eqnarray}
\gamma_{12}(\vec{k})&=&\gamma_{12a}+e^{-i(\vec{a}_{1}+\vec{a}_{2})\cdot \vec{k}}\gamma_{12b}, \\
\gamma_{13}(\vec{k})&=&\gamma_{13a}+e^{-i\vec{a}_{1}\cdot \vec{k}}\gamma_{13b},\\
\gamma_{23}(\vec{k})&=&\gamma_{23a}+e^{-i\vec{a}_{2}\cdot \vec{k}}\gamma_{23b},
\end{eqnarray}
and
\begin{eqnarray}
D_{1}&=&\gamma_{12a}+\gamma_{13a}+\gamma_{12b}+\gamma_{13b}, \\
D_{2}&=&\gamma_{12a}+\gamma_{23a}+\gamma_{12b}+\gamma_{23b}, \\
D_{3}&=&\gamma_{23a}+\gamma_{13a}+\gamma_{23b}+\gamma_{13b}.
\end{eqnarray}
Here, $\gamma_{12\alpha}$, $\gamma_{13\alpha}$, and $\gamma_{23\alpha}$ are defined as
%%%%%%
\begin{eqnarray}
\label{gamma12}
  \gamma_{12\alpha} &=&
  t_{\alpha} \left\{
  (1-\eta_{\alpha})
  \begin{pmatrix}
  1& 0 \\
   0 &1
  \end{pmatrix}
  +
  \eta_{\alpha}
  \begin{pmatrix}
  \frac{1}{4} & \frac{\sqrt{3}}{4} \\
  \frac{\sqrt{3}}{4} & \frac{3}{4}
  \end{pmatrix}
   \right\}, \\
\label{gamma13}
  \gamma_{13\alpha} &=&t_{\alpha} \left\{
  (1-\eta_{\alpha})
  \begin{pmatrix}
   1& 0 \\
   0 &1
  \end{pmatrix}
  +
  \eta_{\alpha}
  \begin{pmatrix}
  1 & 0 \\
  0 & 0
  \end{pmatrix}
 \right\},\\
 \label{gamma23}
  \gamma_{23\alpha} &=&t_{\alpha} \left\{
  (1-\eta_{\alpha})
  \begin{pmatrix}
  1& 0 \\
   0 &1
  \end{pmatrix} \right. \nonumber \\
  &&\hspace{2.3cm} + \left.
  \eta_{\alpha}
  \begin{pmatrix}
  \frac{1}{4} & -\frac{\sqrt{3}}{4} \\
  -\frac{\sqrt{3}}{4} & \frac{3}{4}
  \end{pmatrix}
  \right\}.
\end{eqnarray}
%%%%%%
\end{subequations}
%%%%%%%%%%%%
%}txtb

The Euler-Lagrange equation for $u^{\mu}_{\vec{k},p}$ is written as
%%%%%%%%%%%%
\begin{equation}
\frac{d}{dt} \left(\frac{\partial \mathcal{L}}{\partial \dot{u}^{\mu}_{\vec{k},p} } \right)- \frac{\partial \mathcal{L}}{\partial u^{\mu}_{\vec{k},p} }=0.
\end{equation}
%%%%%%%%%%%%
This leads to an equation of motion under the periodic boundary condition,
%%%%%%%%%%%%
\begin{equation}
 \ddot{u}^{\mu}_{\vec{k},p} + \sum_{q} \Gamma^{\mu\nu}_{pq}(\vec{k}) u_{\vec{k},q}^{\nu}=0.
\end{equation}
%%%%%%%%%%%%
Writing the time independence of $u_{\vec{k},p}$ as
%%%%%%%%%%%%
\begin{equation}
 u^{\mu}_{\vec{k},p}= e^{-i\omega t} \phi^{\mu}_{p}(\vec{k}),
\end{equation}
%%%%%%%%%%%%
we obtain the Euler-Lagrange equation reduced to the eigenvalue equation
%%%%%%%%%%%%
\begin{equation}
\label{eigen_eq_periodic}
 -\omega^{2} \phi_{p}^{\mu}(\vec{k})+\sum_{q} \Gamma_{pq}^{\mu\nu}(\vec{k})\phi_{q}^{\nu}(\vec{k} ) =0.
\end{equation}
%%%%%%%%%%%%
By solving the eigenvalue equation (\ref{eigen_eq_periodic}), we obtain the dispersion relation, which we will discuss in the next section.

Before closing this section, we address the correspondence between the spring-mass model and the tight-binding model.
In fact, the spring-mass model is reduced to two copies of the tight-binding model
if we set $\eta_{a}=0$, $\eta_{b}=0$, i.e., the tension is infinitely strong~\cite{footnote_one},
since the off-diagonal parts of the matrix $\gamma_{ij\alpha}$ vanish [see Eqs. (\ref{gamma12})-(\ref{gamma23})].

%\input{bulk.tex}
%Bund structure under the periodic boundary condition and bulk topological properties Z3 Berry phase up to two bands
%-----The bulk dynamical matrix of spring mass models on a breathing kagome lattice

%The quantization of Z3 Berry phase
\section{$\mathbb{Z}_3$ Berry phase\label{three}}
\subsection{$\mathbb{Z}_3$ Berry phase in a momentum space \label{three-one}}
\begin{figure*}[htp]
  \includegraphics[width=0.8\linewidth]{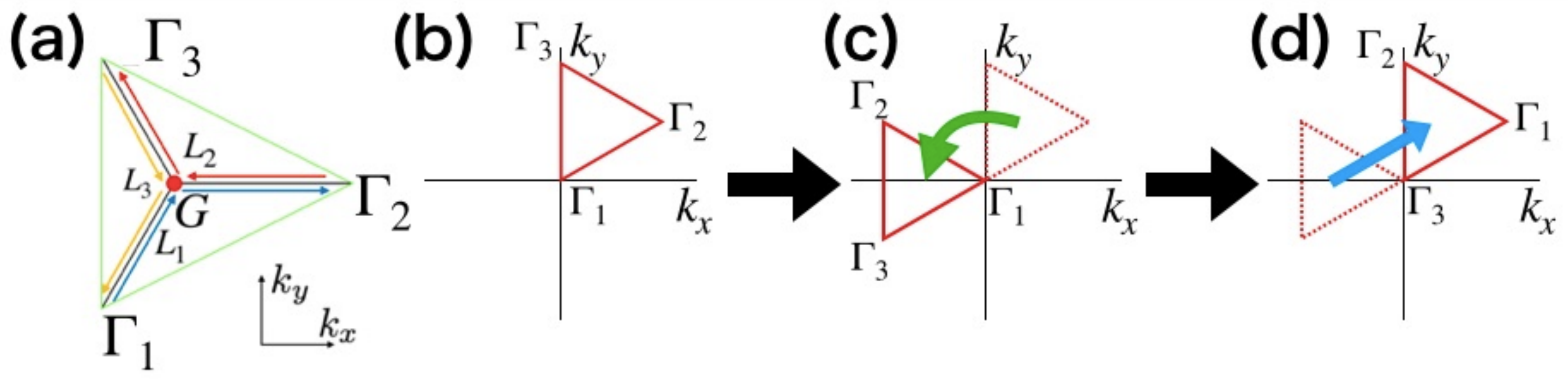}
\caption{(Color online)
 (a) The paths in momentum space. The coordinates of $\Gamma_{1}$, $\Gamma_{2}$, and $\Gamma_{3}$ are
$(0,0)$, $\frac{2\pi}{R_{a}+R_{b}}(1,\frac{1}{\sqrt{3}})$, and $\frac{2\pi}{R_{a}+R_{b}}(0,\frac{2}{\sqrt{3}})$, respectively.
The $G$ point is $\frac{2\pi}{R_{a}+R_{b}}(\frac{1}{3},\frac{1}{\sqrt{3}})$.
The path $L_{i}$ is $\Gamma_{i} \rightarrow G \rightarrow \Gamma_{i+1}\ (\Gamma_{4}=\Gamma_{1})$.
(b) The triangle area whose vertices are $\Gamma_{1}$, $\Gamma_{2}$, $\Gamma_{3}$.
(c) and (d) Schematic figures for the operations which keep the Hamiltonian invariant.
(c) The $120^{\circ}$ rotation of the triangle area in a momentum space.
(d) The translation of the triangle area in a momentum space.
Combining these two operations, the $\Gamma_{i}$ point is transformed to $\Gamma_{i+1}$ with $\Gamma_{4}:=\Gamma_{1}$.
Consequently, the path $L_{i}$ is transformed to $L_{i+1}$ with $L_{4}=L_{1}$.}
\label{rotation}
\end{figure*}

In this section, we introduce the $\mathbb{Z}_3$ Berry phase defined in momentum space.
The key idea originates from the quantized Berry phase with respect to the local twists of the Hamiltonian~\cite{PhysRevLett.123.196402, PhysRevResearch.2.012009, Hatsugai_2011,200615354, Hatsugai_2007, PhysRevB.77.094431, Hatsugai_2010, PhysRevE.87.021301, PhysRevB.94.205112, PhysRevLett.120.247202, PhysRevB.98.195127, kawarabayashi2019fractionally, doi:10.7566/JPSJ.88.104703}.
Such a Berry phase has been used
as a topological order parameter for various topological phases, especially in correlated systems such as spin systems~\cite{200615354, Hatsugai_2007, PhysRevB.77.094431, Hatsugai_2010, PhysRevE.87.021301, PhysRevB.94.205112, PhysRevLett.120.247202, PhysRevB.98.195127, kawarabayashi2019fractionally}.
The Berry phase is quantized due to symmetries, e.g.,
time-reversal symmetry, inversion symmetry and discrete rotational symmetry.
Recently, it was also used to characterize the HOTI phases~\cite{PhysRevLett.123.196402,PhysRevResearch.2.012009,doi:10.7566/JPSJ.88.104703}.

Here, we show that such a Berry phase can also be defined in the momentum-space representation
for the breathing-kagome-lattice spring-mass model.
Extension of the $\mathbb{Z}_3$ Berry phase in momentum space is important from the viewpoint of computational costs.
Namely, to calculate the quantized Berry phase with respect to the local twists of the Hamiltonian,
one has to calculate the many-body ground state under the local twist.
On the other hand, the single-particle eigenfunctions are enough to calculate
the $\mathbb{Z}_3$ Berry phase in momentum space, thus, we can save computational costs when dealing
with non-interacting quantum systems and classical systems.

The $\mathbb{Z}_{3}$ Berry phase for the lowest $\nu$th bands $\gamma^{\nu}$ is defined as follows.
First, we define the $\nu \times \nu$ Berry connection matrix:
\begin{equation}
\vec{A}^{\nu}(\vec{k})= i  \Phi^{\nu\dagger}(\vec{k}) \frac{\partial}{\partial \vec{k}} \Phi^{\nu}(\vec{k}),
\end{equation}
where
\begin{equation}
\begin{split}
\Phi^{\nu}(\vec{k}) &= \left[
\begin{array}{ccc}
\vec{\phi}_{1}(\vec{k}), &
\cdots &
,\vec{\phi}_{\nu}(\vec{k}) \\
\end{array}
\right] \\
&=
\begin{pmatrix}
 (\phi_{1})_1 & (\phi_{2})_1 & \cdots & (\phi_{\nu})_1 \\
 \vdots & \vdots & \ddots & \vdots \\
 (\phi_{1})_6 & (\phi_{2})_6 & \cdots & (\phi_{\nu})_6
\end{pmatrix},
\end{split}
\end{equation}
is the $6 \times \nu$ matrix composed of
the eigenvectors of a momentum-space dynamical matrix in the spring-mass model on the breathing kagome lattice,
represented by $\vec{\phi}_{n}(\vec{k})$.
Then, the Berry phase for the lowest $\nu$ bands is expressed as
\begin{equation}
 \gamma^{\nu}(L_i)=\int_{L_{i}}  \mathrm{Tr} \ [\vec{A}^{\nu}(\vec{k}) ]\cdot  d\vec{k}.
 \label{zberry}
\end{equation}
where $L_{i}(i=1,2,3)$ is a path in momentum space, $L_{i}:\Gamma_{i} \rightarrow G \rightarrow \Gamma_{i+1}(\Gamma_4=\Gamma_1)$ [Fig.~\ref{rotation}(a)].

In the spring-mass model on a breathing kagome lattice, the moemntum-space dynamical matrix is invariant under the three-fold rotation in momentum space (Fig.~\ref{rotation}).
We define this operator as $U$.
The momentum-space dynamical matrix has a symmetry which is expressed as
\begin{equation}
 U\Gamma(k_{i})U^{-1} =\Gamma(C_3k_i),
\end{equation}
with $k_i \in L_{i}$ and $L_{i}$ being a path in a momentum space, $L_{i}:\Gamma_{i} \rightarrow G \rightarrow \Gamma_{i+1}$.
Here, we have supposed that the momentum $k_i$ is mapped to $C_3k_i\in L_{i+1}$ by applying $C_3$ rotation.
This relation indicates that the Berry phases $\gamma^{\nu}(L_{i})$ computed along each path take the same value,
\begin{equation}
 \gamma^\nu(L_{1})= \gamma^\nu(L_{2})= \gamma^\nu(L_{3}). \label{eq:gamma_threepaths}
\end{equation}
%%%%%%
In addition, the integral along the path $L_{1} + L_{2}+ L_{3}$ is equal to zero,
%%%%%%
\begin{equation}
 \label{sum}
 \sum_{i=1}^{3} \gamma^\nu(L_{i})=0 \quad  \mathrm{mod} \ 2 \pi.
\end{equation}
From Eqs.~(\ref{eq:gamma_threepaths})~and~(\ref{sum}),
\begin{equation}
 \gamma^\nu \equiv  \gamma^\nu(L_{i}) =\frac{2 \pi k}{3} \quad \mathrm{mod} \ 2\pi,
\end{equation}
where $k$ is $0$, $1$, or $2$.
The same argument can be applied to a tight-binding model on a breathing kagome lattice
by replacing the momentum-space dynamical matrix $\Gamma$ with the Hamiltonian $H$
since $H$ preserves three fold rotational symmetry and translational symmetry.
Note that, in the spring-mass models, the rotation is applied not only to the momentum and sublattice degrees of freedom
but also to directions of the displacement ($\mu=x,y$).

In fact, the $\mathbb{Z}_3$ Berry phase in a momentum space is equivalent to the local-twist Berry phase discussed in Ref.~\onlinecite{Hatsugai_2011}.
To see this, we consider the $\mathbb{Z}_3$ Berry phase in a momentum space, where the upward triangle is a unit cell.
In this case, there is the phase factor from the Bloch wave vector for hoppings from a site in the upward triangle to a site in
the downward triangle.
By replacing the factors  $e^{- i \vec{k}\cdot \vec{a}_1}$ and $e^{-i \vec{k}\cdot \vec{a}_2}$ with the
twisting parameters $e^{i \theta_1}$ and $e^{i \theta_2}$ respectively,
we find that the $\mathbb{Z}_3$ Berry phase in a momentum space is same as the local-twist Berry phase for the $1\times 1$ unit cell
of the downward triangle.
A similar correspondence also holds for the Su-Schrieffer-Heeger model~\cite{PhysRevLett.42.1698} and the breathing pyrochlore model.

%%%%%%%%%%%%
\begin{figure*}[ht]
 \centering
  \includegraphics[width=0.9\linewidth]{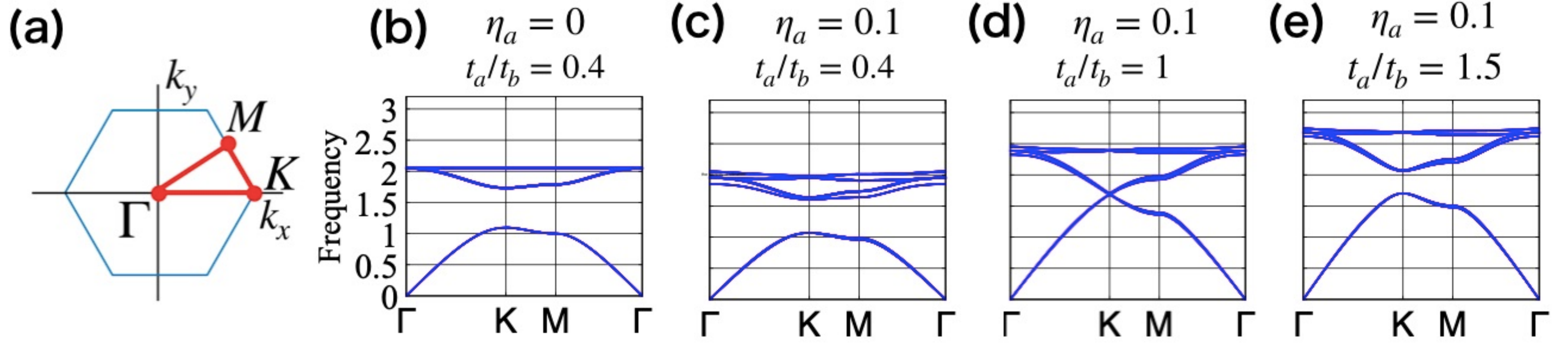}
 \caption{
(Color online)
Bulk properties of the spring-mass model.
(a) The first Brillouin zone of the spring-mass model. The $\Gamma$, $K$, and $M$ points are located at $(k_x,k_y)=(0,0)$, $\frac{2\pi}{R_a+R_b}(\frac{2}{3},0)$, and $\frac{2\pi}{R_a+R_b}(\frac{1}{2},\frac{1}{2\sqrt{3}})$, respectively.
(b)-(e) The band structure along the line shown in (a) for several parameter sets $(\eta_\alpha, t_a/t_b)$. (b), (c), (d), and (e) show the data for $(\eta_\alpha, t_a/t_b)=(0,0.4)$, $(0.1,0.4)$, $(0.1,1)$, and (0.1,1.5), respectively.
}
 \label{band1}
\end{figure*}
%%%%%%%%%%%%

\subsection{Bulk properties\label{three-rwo}}
In this section, we present the properties of this model under the periodic boundary condition.
To be specific, we investigate the dispersion relation and the bulk topological invariant.
First, we show the band structures obtained by diagonalizing the momentum-space dynamical matrix [Fig.~\ref{band1}(b)-(e)].
Here, the horizontal axis denotes the high-symmetry lines in the Brillouin zone with
$\Gamma$, $K$, and $M$ denoting the high-symmetry points [Fig.~\ref{band1}(a)].

There are six bands; for $\eta_{a}=0$, we see three bands, each of which is doubly degenerate.
For $\eta_{a} \neq 0$, two fold degeneracy is lifted because of the coupling
between transverse
waves and longitudinal waves [Figs.~\ref{band1}(b)~and~\ref{band1}(c)].
Moreover, there are no flat bands unless $\eta_{a}= 0$.
These are the unique characters of the spring-mass model which are different from the tight-binding model.

\begin{figure}[htp]
  \includegraphics[width=\linewidth]{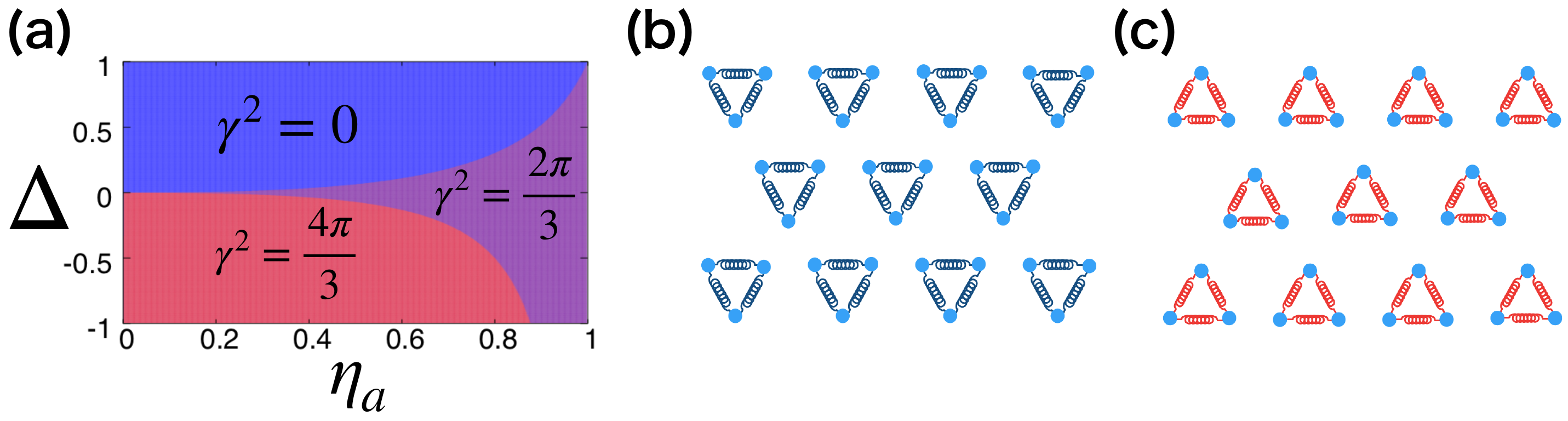}
\caption{(Color online)
(a) The Berry phase for the lowest two bands. (b)-(c) Schematic figures for two limits with (b) $t_a=0$ and (c) $t_b = 0$.
}
\label{correspondence}
\end{figure}

In Fig.~\ref{correspondence}(a), we show the numerical results for the $\mathbb{Z}_{3}$ Berry phase with $\nu=2$.
For the numerical calculation, we employ the method introduced in Ref.~\onlinecite{fukui2005chern} to avoid the difficulties in the gauge choice.
Note that the $\mathbb{Z}_{3}$ Berry phase cannot be defined for $\eta_{a}=1$ because the gap between the lowest second and third bands is closed.
To map out the $\mathbb{Z}_{3}$ Berry phase in the parameter space, we introduce
$\Delta \in [-1,1]$, describing the degrees of breathing.
More precisely, $\Delta$ relates $t_a$ and $t_b$ as
\begin{subequations}
\begin{equation}
t_a =  1 + \Delta
\end{equation}
and
\begin{equation}
t_b = 1 - \Delta.
\end{equation}
\end{subequations}
For $\Delta = 1$ ($-1$), the system is reduced to a set of isolated triangles of mass points connected by red (blue) springs. We note that $t_b = 0$ ($t_a = 0$) holds for $\Delta = 1$ ($-1$).
For $\Delta = 0$, the mass points form an isotropic kagome lattice because $t_a = t_b$ holds.

We see in Fig.~\ref{correspondence}(a) that there are three phases where the $\mathbb{Z}_3$ Berry phase takes $\gamma=0, \frac{2\pi}{3}$, and $\frac{4\pi}{3}$, respectively.
Among them, the phase with $\gamma = \frac{2\pi}{3}$ is the new phase which does not have a counterpart in the tight-binding model.
In fact, this phase originates from the interaction between the longitudinal and transverse waves inherent in the spring-mass model.
%----------------------%
%---------------------%
We can observe the gap closing between the second and third bands when the $\mathbb{Z}_{3}$ Berry phase changes [see Fig.~\ref{band1}(c)-\ref{band1}(e)].
In other words, the $\mathbb{Z}_{3}$ Berry phase is
the topological invariant, so its value does not change
as long as the band gap is not closed.
This indicates that the system can be adiabatically deformed to a certain limit that has the same Berry phase.
In the present model, for small $\eta_a$, the phase with $\gamma = \frac{4 \pi}{3}$ is connected to the limit of $\Delta = -1$ (i.e., $t_a = 0$),
while the phase with $\gamma = 0$ is connected to the limit of $\Delta = 1$ (i.e., $t_b = 0$).
The schematic figures for the former and the latter are illustrated in Figs.~\ref{correspondence}(b)~and~\ref{correspondence}(c), respectively.
This physical picture of the adiabatic connection to the decoupled triangles is essential to
understand the bulk-corner correspondence in this system, as we will explain in the next section.

%\input{fixed.tex}
%Bund structure under the fixed boundary condition and the existence of the corner states
\section{Corner states under the fixed boundary condition\label{four}}
\begin{figure*}[ht]
 \includegraphics[width=\linewidth]{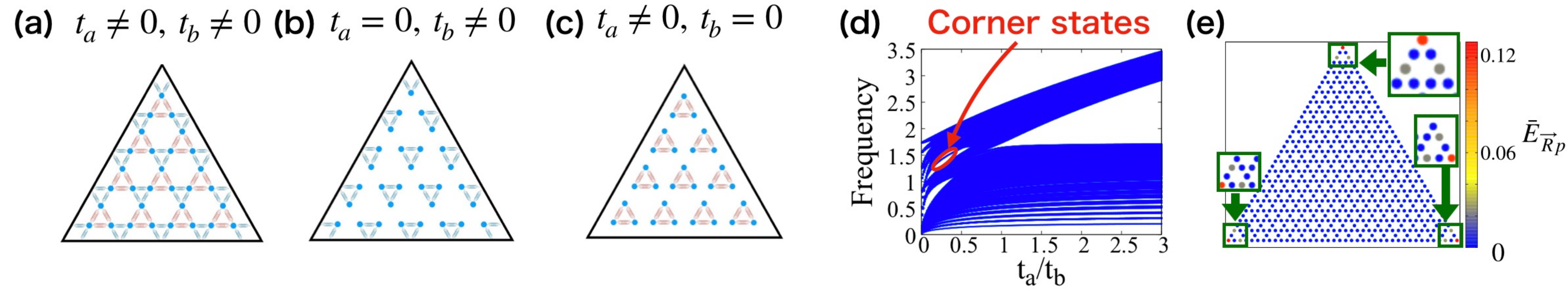}
 \caption{(Color online)
Schematic figures for
the spring-mass model in a triangle arrangement for
 (a)$t_{a}\neq 0, t_{b}\neq 0$, (b) $t_{a}= 0, t_{b}\neq 0$, and (c) $t_{a}\neq 0, t_{b}= 0$.
 Black lines represent walls.
(d) Eigenfrequencies as a function of $t_{a}/t_{b}$ for $\eta_{a}=0.1$ with a triangle arrangement.
(e) Kinetic-energy distribution in the real space of the corner states for $t_{a}/t_{b}=0.4$.
We write the number of small triangles along the edge as $L$; the total number of masses is $3L(L+1)/2$.}
 \label{corner_fig}
\end{figure*}

In this section, we demonstrate that system hosts corner states due to topological properties in the bulk.
This bulk-corner correspondence serves as a direct evidence of the higher-order topological phase.
Specifically, considering a triangle arrangement~\cite{PhysRevLett.120.026801} (Fig.~\ref{corner_fig}),
we numerically show that the corner states emerge for $\gamma^2=4\pi/3$ while they do not for $\gamma^2=0$.

Before going to the numerical results, we consider two limits, i.e., $t_{a}=0$ [Fig.~\ref{corner_fig}(b)], and $t_{b}=0$ [Fig.~\ref{corner_fig}(c)],
to gain insight into the boundary states.
Note that, in these limits, the equilibrium condition of Eq.~(\ref{equiv}) is inevitably broken.
Nevertheless, it is helpful to consider these limits, as we explain below.
For $t_{a}= 0,\ t_{b} \neq 0$ [Fig.~\ref{corner_fig}(b)], there exist three isolated mass points connected only to the wall,
at three corners of the triangle.
This configuration supports the eigenmodes localized at the corners.
In contrast to this, there are no isolated mass points for $t_{a}\neq 0,\ t_{b} = 0$ [Fig.~\ref{corner_fig}(c)].
From these, we expect that three corner states exist for $t_a \ll t_b$, while they do not for $t_b \ll t_a$.

Keeping this observation in mind, let us move on to the numerical results.
In Fig.~\ref{corner_fig}(d), we plot the energy spectra as a function of
$t_{a}/t_{b}$ for $\eta_{a}=0.1$ and $L=20$
(see Appendix~\ref{eight} for the result for large $\eta_a$).
We see the existence of the in-gap states for a certain region of $t_a /t_b < 1$, encircled by a red ellipse in Fig.~\ref{corner_fig}(d).
Note that these corner states have quasithreefold degeneracy for $\eta_{a} \neq 0$.
We also note that, even for $t_a /t_b < 1$, the corner states may be energetically buried in the bulk or edge states; thus, they cannot be seen in the energy spectra of Fig.~\ref{corner_fig}(d);
for the edge state of the present model, see Appendix~\ref{nine}.
To avoid this problem, we calculate the IPR and show the corner states in the bulk or edge continuum (see Appendix~\ref{ten}).
Figure~\ref{corner_fig}(e) indicates that the in-gap states observed above corresponds to the corner states.
This figure shows the kinetic-energy distribution
\begin{equation}
 \bar{E}_{\vec{R},p}=\frac{1}{4N_{\rm c}}\sum_{\ell=1}^{N_{\rm c}} \left( \omega_{\ell} I^{\ell}_{\vec{R},p}\right)^{2} \label{eq:kin}
\end{equation}
for these quasi-degenerate in-gap states.
Here, $I^{\ell}_{\vec{R},p} = \sqrt{ (\xi^{\ell,x}_{\vec{R},p})^2 + (\xi^{\ell,y}_{\vec{R},p})^2 } $ is the amplitude of the displacement of the mass point $(\vec{R},p)$ of the mode $\ell$. The summation in Eq.~(\ref{eq:kin}) is taken over the $N_{\rm c}$-fold (quasi)degenerate states.
In Fig.~\ref{corner_fig}(e), we can clearly see that the kinetic energy distribution is localized at the corners, manifesting the existence of the higher-order topological phase in the present model.

Combining the above results and the fact that the $\mathbb{Z}_{3}$ Berry phase $\gamma^2$ takes $4\pi/3$ for $t_a /t_b < 1$ (see Fig.~\ref{correspondence}), we can confirm that the bulk-corner correspondence holds for our spring-mass model.
This is a direct consequence of the adiabatic connection argument we have presented in Sec.~\ref{three}.

In the above we have focused on the case where $\eta_a$ is small ($\eta_a =0.1$).
We can also observe the bulk-corner correspondence for $\eta_a =0.9$, where the in-gap states are between the third and the fourth bands (see Appendix~\ref{eight}).
We also note that the corner states are also found in a parallelogram arrangement~\cite{2017arXiv171109202X,PhysRevB.97.241405},
which can also be understood in the same adiabatic connection argument (see Appendix~\ref{eleven}).

%\input{forced.tex}
%Suggestion how to observe the corner state in this model
\section{Forced vibration\label{five}}

\begin{figure}[ht]
 \centering
  \includegraphics[width=\linewidth]{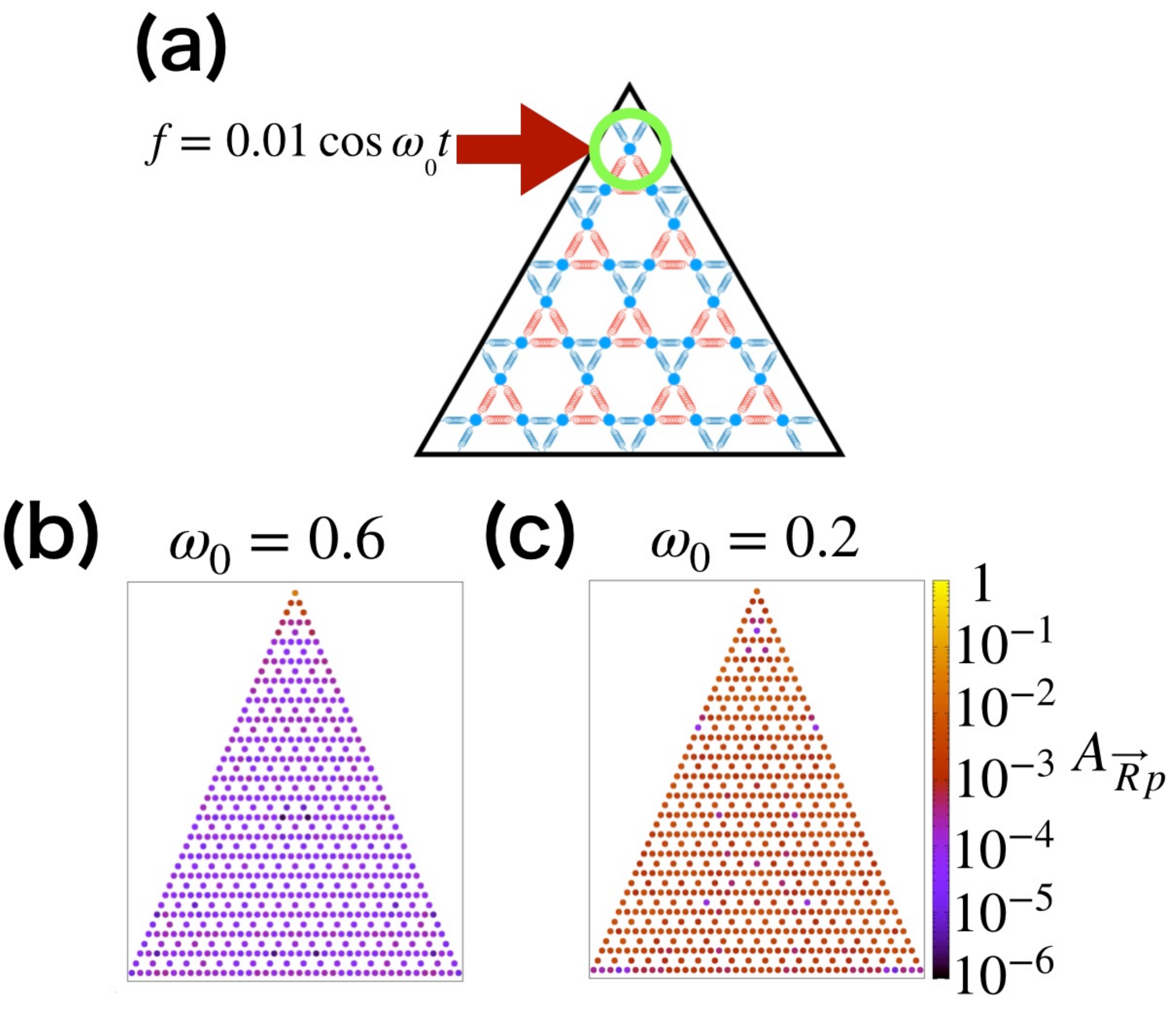}
\caption{(Color online)
(a) Schematic figure for the experimental setup for the forced vibration.
The force $f(t)$ is added to the mass point of the top corner of the triangle (encircled by a green circle).
(b) $\omega_0$ is near a corner mode,
and
(c) $\omega_0$ is away from a corner mode.
In (b) and (c), the amplitudes $A_{\vec{R},p}$ at $t=1000$ are plotted.
}
\label{forced}
\end{figure}

In this section, we point out that corner states emerging from our spring-mass model can be experimentally observed by analyzing a forced vibration.
The forced vibration is caused by an external force $f(t)$.
The equation of motion with an external force is
\begin{equation}
\label{force_eq}
\ddot{\vec{x}}+D \vec{x}=\vec{f}(t),
\end{equation}
where $(f)_{\vec{R},p,\mu}(t)=f^{\mu}_{\vec{R},p}(t)$ is an external force which is added to the mass at $\vec{R} + \vec{r}_p$
along the $\mu$ direction.

Consider the case where the external force is given by $f^{\mu}_{\vec{R},p}(t)=F^{\mu}_{\vec{R}, p} \cos(\omega_{0} t )$;
that is, it oscillates in time with the frequency $\omega_0$.
We further set $F^{\mu}_{\vec{R}, p}$ such that it has amplitudes only at the top corner [Fig.~\ref{forced}(a)].
Then, as a particular solution of the equation of motion of
Eq.~(\ref{force_eq}), $\xi^\prime_i$, which governs the behaviors of the long time scale,
we get
\begin{equation}
\label{forced_solution}
\xi_{i}^\prime
=
\begin{cases}
 \frac{\sum_{j}U_{ij}F_{j}}{\omega_{i}^{2}-\omega_{0}^{2}} \cos \omega_{0} t &\mathrm{for} \quad \omega_{i} \neq \omega_{0}, \\
 \frac{\sum_{j}U_{ij}F_{j}}{2 \omega_{0}}t \sin \omega_{0} t &\mathrm{for} \quad \omega_{i} = \omega_{0},
\end{cases}
\end{equation}
where $U$ is a matrix obtained by aligning the eigenvectors of the equation of motion in the absence of
$f(t)$ [i.e., Eq.~(\ref{eigenreal})],
$i$ specifies the eigenmode,
$j$ is the abbreviation of the position of the mass point $\vec{R} + \vec{r}_p$ and the direction of the motion $\mu$,
and $\omega_i$ is the eigenfrequency of the $i$th mode.
From Eq.~(\ref{forced_solution}),
we see the resonance occurs at $\omega_{i}=\omega_{0}$.
This indicates that for $\omega_0$ being close to the eigenfrequency of the corner state,
one obtains the large vibration amplitude only at the corner;
in contrast, when it is close to the eigenfrequency of the bulk state,
the vibration spreads over the bulk.
Thus, by looking at the time evolution of the forced vibration
upon changing $\omega_0$,
one can determine whether the corner states exist or not.

To demonstrate this,
we numerically solve the equation of motion using the Euler method on a triangle arrangement.
In the Euler method, the time revolution is described as
\begin{equation}
\begin{split}
\left(
 \begin{array}{c}
  \vec{x}(t+\Delta t)\\
  \dot{\vec{x}}(t+\Delta t)
 \end{array}
\right)
&=
\left\{
\begin{pmatrix}
 \hat{0} & \hat{1} \\
 -D & \hat{0}
\end{pmatrix}
\left(
 \begin{array}{c}
  \vec{x}(t)\\
  \dot{\vec{x}}(t)
 \end{array}
\right)
+
\left(
 \begin{array}{c}
  \vec{0} \\
  \vec{f}(t)
 \end{array}
\right)
\right\}
\Delta t
\\
& \quad+
\left(
 \begin{array}{c}
  \vec{x}(t)\\
  \dot{\vec{x}}(t)
 \end{array}
\right),
\end{split}
\end{equation}
where
$\Delta t$ is a small time step.
$D$ is given in Eq.~(\ref{force_eq}).
For the numerical simulations, we set $\eta_{a}=0.9$, $t_{a}/t_{b}=0.1$, and $\Delta t=0.0001$.
It is worth noting that $\eta$ is close to 1 for realistic springs.
The initial state is set as $x^{\mu}_{\vec{R},p}(0)=0$ and $\dot{x}^{\mu}_{\vec{R},p}(0)=0$; that is, the system is in equilibrium.
The external force in the $x$ direction is added to the mass point at the top of the corner; we could not observe the qualitative difference when the external force is in the $y$ direction.
After running a simulation to $t = t_{\rm max}=1000$,
we observe the amplitudes of vibration,
\begin{equation}
 A_{\vec{R},p}=\sqrt{[x^{x}_{\vec{R},p} (t_{\rm max})]^{2}+[x^{y}_{\vec{R},p} (t_{\rm max})]^{2}}.
\end{equation}

The results are shown in Figs.~\ref{forced}(b)~and~\ref{forced}(c).
As expected, when $\omega_{0}$ is near the frequency of the corner state~\cite{footnote_two}, the large vibration is seen only near the corner,
while the vibration propagates in the bulk when $\omega_{0}$ is near the frequency of the bulk state.

The above results suggest that the corner state induced by the topological properties in the bulk can be experimentally observed; the resonance frequency corresponds to the frequency of corner states.

%\input{summury.tex}

%Summary
\section{summary\label{six}}
In summary, we have shown that the higher-order topological phase is realized in the spring-mass model on a breathing kagome lattice.
We have introduced the $\mathbb{Z}_3$ Berry phase in the Brillouin zone and found that our bulk topological invariant characterizes the topologically nontrivial phase.
Remarkably, we have found the topologically nontrivial phase with $\gamma = \frac{2\pi}{3}$, in addition to
the phase with $\gamma = \frac{4\pi}{3}$ that corresponds to the two copies of the topological phase in the tight-binding model.
This is due to the coupling between longitudinal and transverse modes, inherent in the spring-mass model.
We have also found that the characteristic corner states appear under the fixed boundary condition
in both the triangle and parallelogram arrangements.
In addition, we have proposed that the corner states can be detected experimentally
through a forced vibration.
By the numerical simulation, we have found that the corner-selective vibration is observed when the external frequency is close to that of the corner modes.

\acknowledgements
This work is partly supported by JSPS KAKENHI Grants No.~JP16K13845, No.~JP17H06138, No.~JP18H05842, No.~JP19J12315, and JP19K21032.
Part of numerical calculations were performed on the supercomputer at ISSP at the University of Tokyo.

\appendix
%\input{fifth.tex}
% Z3 Berry phase up to five bands
\section{\texorpdfstring{$\mathbb{Z}_{3}$}{Lg} Berry phase for the lowest five bands\label{eight}}
\begin{figure}[htb]
 \centering
  \includegraphics[width=\linewidth]{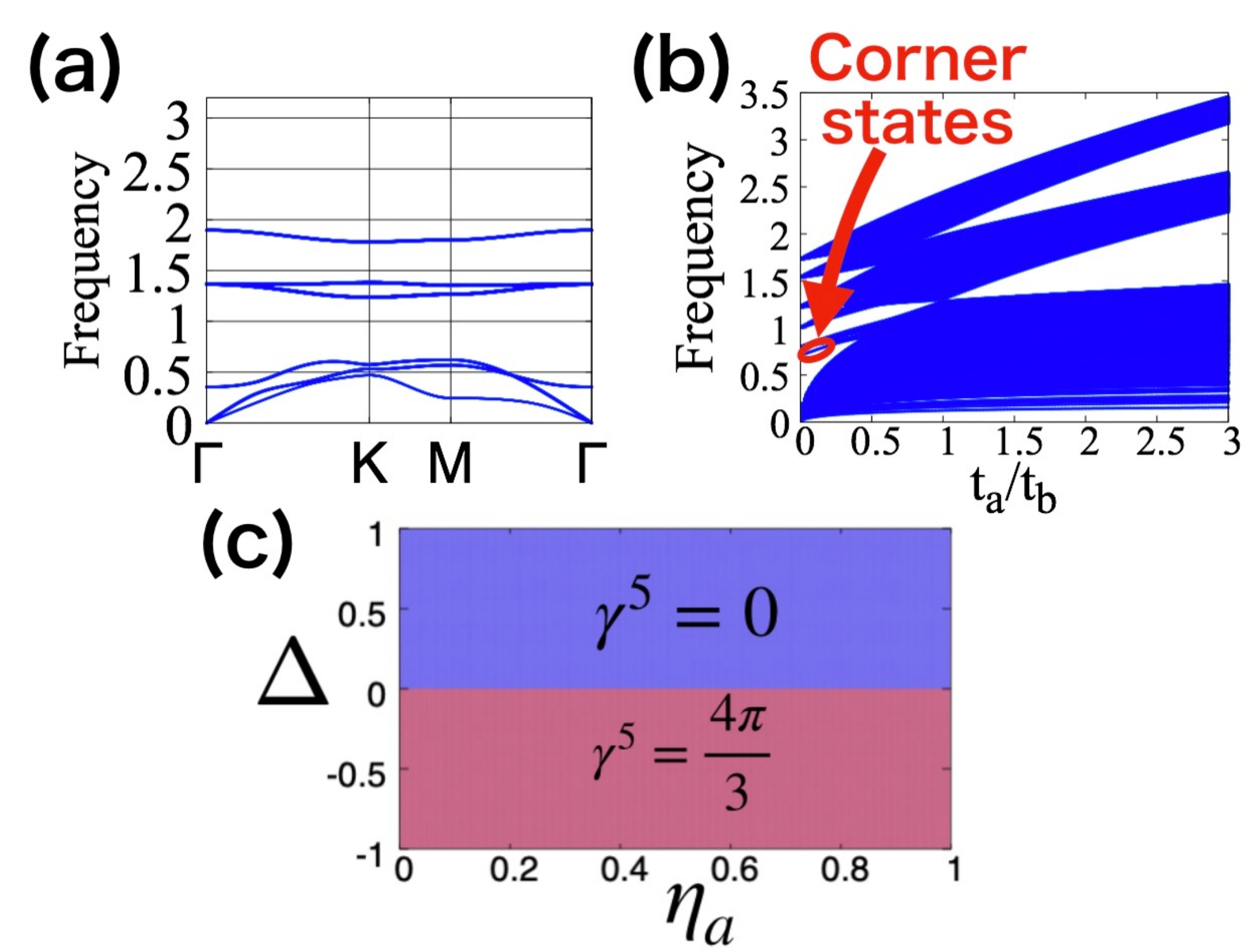}
\caption{(Color online)
(a) The bands structure for $\eta_{a}=0.9$, $t_{a}/t_{b}=0.1$.
(b) The energy spectrum for $\eta_{a}=0.9$.
The corner modes are between third and fourth bands.
(c) The $\mathbb{Z}_{3}$ Berry phase for the lowest fifth bands.
}
\label{berry_fifth}
\end{figure}

In this appendix, we show the result for
$\gamma^{5}$, i.e., the Berry phase for the lowest five bands.
The reason for calculating $\gamma^{5}$ is that
there exist corner states that
appear between the third and fourth bands for $\eta_{a} \sim 1$, contrary to the case of $\eta_{a} \sim 0$.
To be concrete, we show the bulk band structure and the energy spectra
in a triangle arrangement at $\eta_{a} = 0.9$ in Figs.~\ref{berry_fifth}(a)~and~\ref{berry_fifth}(b), respectively, and the $\mathbb{Z}_3$ Berry phase for the lowest five bands.

We find that the band gap exists between the third and fourth bands
and between fifth and the sixth bands, in contrast to the case of $\eta_{a} \sim 0$.
Correspondingly, the corner states appear for a certain region for $t_a < t_b$, while there are no corner states for
$t_a > t_b$, which is again inferred from the adiabatic connection argument.

We plot $\gamma^{5}$ in Fig.~\ref{berry_fifth}(c),
showing that $\mathbb{Z}_{3}$ is independent of $\eta_{a}$ and the phase translation occurs at $\Delta=0$, i.e., $t_{a}=t_{b}$.
Similar to $\gamma^{2}$ discussed in Sec.~\ref{three},
$\gamma^{5} = \frac{4\pi}{3}$ indicates that the system is adiabatically connected to the decoupled triangles with blue springs [Fig.~\ref{correspondence}(b)],
while $\gamma^{5} = 0$ indicates that the system is adiabatically connected to the decoupled triangles with red springs [Fig.~\ref{correspondence}(c)].

%\input{cylinder.tex}

%The cylinder of this spring-mass model
\section{The cylinder of the spring-mass model\label{nine}}

\begin{figure}[htp]
  \includegraphics[width=\linewidth]{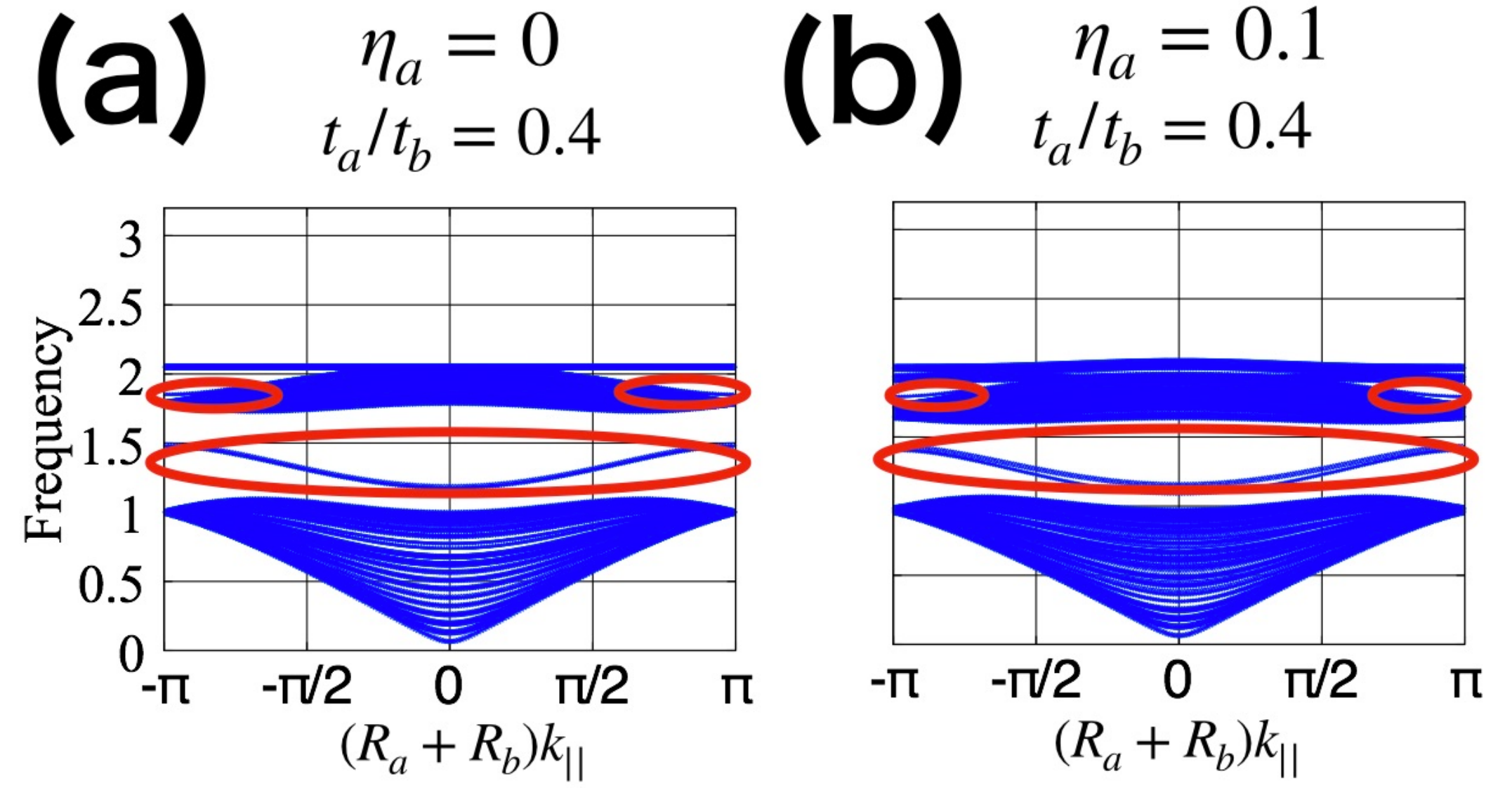}
\caption{(Color online)
The dispersion relations in the cylinder for (a) $\eta_{a} = 0$, $t_{a}/t_{b} = 0.4$ and (b) $\eta_{a} = 0.1$, $ t_{a}/t_{b} = 0.4$.
There exist edge modes between the bulk continua (encircled by red ellipses).
}
 \label{nanoribbons}
\end{figure}

In this appendix, we show the dispersion relations on
the cylinder geometry, focusing on the features of the edge states.
Here, we set the number of red springs in the axial direction as 20,
and we write the momentum in the azimuth direction $k_{||}$.

The results are shown in Fig.~\ref{nanoribbons}.
We see that there exists edge modes between the bulk continua, whose real-space distributions are localized at the edge.
Importantly, the edge mode is not energetically connected to the bulk continuum, meaning that
the lower-dimensional boundary states, i.e., the corner states, are allowed to exist between the edge modes and the bulk modes.

\section{Inverse participation ratio\label{ten}}
%%%%%%%%%%%%
\begin{figure}[H]
 \centering
  \includegraphics[width=0.76\linewidth]{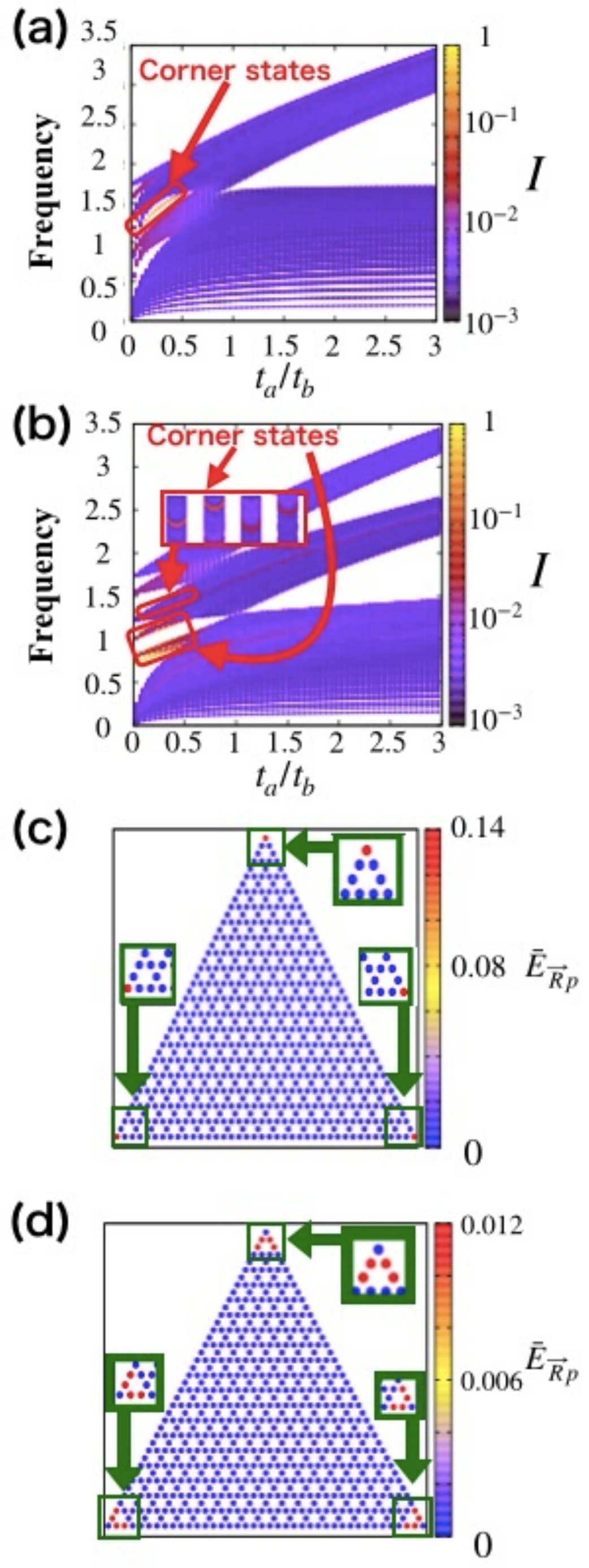}
\caption{(Color online)
(a)-(b) Color map of IPR in the figure of eigenfrequencies.
(a) and (b) show the data for $\eta_a = 0.1$ and $0.9$, respectively.
Kinetic-energy distribution in the real space of the corner states for (c) $t_{a}/t_{b}=0.2$, $\eta_{a}=0.9$, $\omega=1.3954$, and (d) $t_{a}/t_{b}=0.2$, $\eta_{a}=0.9$, $\omega=0.854085$.}
 \label{IPR_fig}
\end{figure}
%%%%%%%%%%%%

%\input{IPR.tex}
\begin{figure*}[t]
 \centering
  \includegraphics[width=\linewidth]{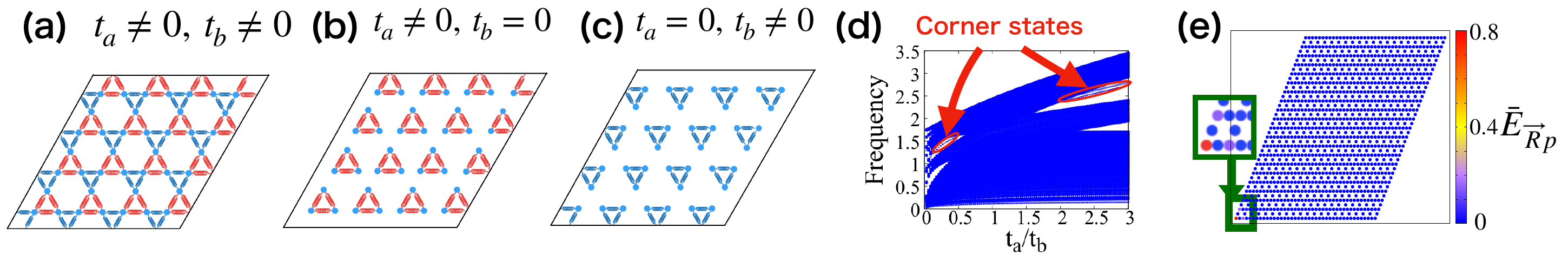}
\caption{
Schematic figures for the spring-mass model in a parallelogram arrangement with
(a) $t_{a}\neq 0,\ t_{b} \neq 0$, and (c) $t_{a}\neq 0,\ t_{b}=0$, and (c) $t_{a}=0,\ t_{b}\neq 0$.
Black lines represent walls.
(d) Eigenfrequencies as a function of $t_{a}/t_{b}$ for $\eta_{a}=0.1$ with a parallelogram arrangement.
(e) Kinetic-energy distribution in the real space of the corner states for $t_{a}/t_{b}=0.4$.
The corner mode for $\gamma=0$ arises from the fact that we have cut the unit cells which make the bulk-edge correspondence ubiquitous.
}
\label{corner_fig_para}
\end{figure*}
In this section, we show the numerical result of the IPR to specify the corner states buried in the bulk or edge continuum.
In the literature, the IPR has been used to study Anderson localization in disordered systems~\cite{doi:10.1143/JPSJ.43.415,JANSSEN19981},
and it was recently used to specify the corner states in the HOTI~\cite{PhysRevB.99.085406}.
The IPR in the spring-mass model is defined as
\begin{equation}
 \label{IPR}
 I=\sum_{\vec{R}}\sum_{p}\left\{\sum_{\mu}(\xi^{\mu}_{\vec{R},p})^2\right\}^2,
\end{equation}
where $\vec{\xi}$ stands for the normalized eigenvector.
For the extended states, the relation $\sum_{\mu}(\xi^{\mu}_{\vec{R},p})^2 \simeq 1/N$ holds, which results in $I \simeq 1/N $.
Therefore, the IPR for the extended states vanishes in the large-system-size limit.
In contrast, for the corner states, the relation $\sum_{\mu}(\xi^{\mu}_{\vec{R},p})^2 \simeq \delta_{\vec{R},\vec{R}_{c}} \delta_{p,p_{c}}$ holds,
which results in $I \simeq 1$.
Here, $(\vec{R}_{c},\ p_{c})$ denotes the corner site.

The numerical result is shown in Fig.~\ref{IPR_fig}.
We see that the corner modes in the bulk or edge continuum are specified by the large IPR [Fig.~\ref{IPR_fig}(c)].

Additionally, we find the ``cornerlike'' modes,
where the eigenvector has large amplitudes not right at the corners but at the sites near the corners [Fig.~\ref{IPR_fig}(d)].
Such modes appear for large $\eta$ ($\eta = 0.9$ in this case);
thus, they are characteristic of the spring-mass model.

\section{Parallelogram arrangement under the fixed boundary condition\label{eleven}}
In this appendix, we show the results for a parallelogram arrangement [Fig.~\ref{corner_fig_para}(a)].
Specifically, we show the corner modes in the parallelogram arrangement and their bulk-corner correspondence.

In this arrangement, we write the number of small triangles along the edge as $L$.
Then, the number of total masses is $3L^{2}+4L+1$.

In Fig.~\ref{corner_fig_para}(d), we plot the energy spectra as a function of $t_{a}/t_{b}$ for $L=20$ and $\eta_{a}=0.1$.
As expected, we see the in-gap states for both $t_a < t_b$ and $t_a > t_b$.
Looking at the spacial distribution of the kinetic energy of the in-gap states
for $t_a < t_b$,
we find that it is localized at the left-bottom corner [Fig.~\ref{corner_fig_para}(e)].

The bulk-corner correspondence in the is explained as follows.
As we saw in Sec.~\ref{three},
$\gamma^2=4\pi/3$ indicates that the band structure is adiabatically connected to that for $t_a=0$
[Fig.~\ref{correspondence}(b)]; similarly, $\gamma^2=0$ indicates that the band structure is adiabatically connected to that for $t_b=0$ [Fig.~\ref{correspondence}(c)].
This means that the former case has the corner state at the top right corner [Fig.~\ref{corner_fig_para}(b)] and the latter case
has the corner state at the bottom  left corner [Fig.~\ref{corner_fig_para}(c)].

%\bibliographystyle{apsrev4-1}
%\bibliography{reference}

\begin{thebibliography}{60}%
\makeatletter
\providecommand \@ifxundefined [1]{%
 \@ifx{#1\undefined}
}%
\providecommand \@ifnum [1]{%
 \ifnum #1\expandafter \@firstoftwo
 \else \expandafter \@secondoftwo
 \fi
}%
\providecommand \@ifx [1]{%
 \ifx #1\expandafter \@firstoftwo
 \else \expandafter \@secondoftwo
 \fi
}%
\providecommand \natexlab [1]{#1}%
\providecommand \enquote  [1]{``#1''}%
\providecommand \bibnamefont  [1]{#1}%
\providecommand \bibfnamefont [1]{#1}%
\providecommand \citenamefont [1]{#1}%
\providecommand \href@noop [0]{\@secondoftwo}%
\providecommand \href [0]{\begingroup \@sanitize@url \@href}%
\providecommand \@href[1]{\@@startlink{#1}\@@href}%
\providecommand \@@href[1]{\endgroup#1\@@endlink}%
\providecommand \@sanitize@url [0]{\catcode `\\12\catcode `\$12\catcode
  `\&12\catcode `\#12\catcode `\^12\catcode `\_12\catcode `\%12\relax}%
\providecommand \@@startlink[1]{}%
\providecommand \@@endlink[0]{}%
\providecommand \url  [0]{\begingroup\@sanitize@url \@url }%
\providecommand \@url [1]{\endgroup\@href {#1}{\urlprefix }}%
\providecommand \urlprefix  [0]{URL }%
\providecommand \Eprint [0]{\href }%
\providecommand \doibase [0]{http://dx.doi.org/}%
\providecommand \selectlanguage [0]{\@gobble}%
\providecommand \bibinfo  [0]{\@secondoftwo}%
\providecommand \bibfield  [0]{\@secondoftwo}%
\providecommand \translation [1]{[#1]}%
\providecommand \BibitemOpen [0]{}%
\providecommand \bibitemStop [0]{}%
\providecommand \bibitemNoStop [0]{.\EOS\space}%
\providecommand \EOS [0]{\spacefactor3000\relax}%
\providecommand \BibitemShut  [1]{\csname bibitem#1\endcsname}%
\let\auto@bib@innerbib\@empty
%</preamble>
\bibitem [{\citenamefont {Hasan}\ and\ \citenamefont
  {Kane}(2010)}]{RevModPhys.82.3045}%
  \BibitemOpen
  \bibfield  {author} {\bibinfo {author} {\bibfnamefont {M.~Z.}\ \bibnamefont
  {Hasan}}\ and\ \bibinfo {author} {\bibfnamefont {C.~L.}\ \bibnamefont
  {Kane}},\ }\href {\doibase 10.1103/RevModPhys.82.3045} {\bibfield  {journal}
  {\bibinfo  {journal} {Rev. Mod. Phys.}\ }\textbf {\bibinfo {volume} {82}},\
  \bibinfo {pages} {3045} (\bibinfo {year} {2010})}\BibitemShut {NoStop}%
\bibitem [{\citenamefont {Qi}\ and\ \citenamefont
  {Zhang}(2011)}]{RevModPhys.83.1057}%
  \BibitemOpen
  \bibfield  {author} {\bibinfo {author} {\bibfnamefont {X.-L.}\ \bibnamefont
  {Qi}}\ and\ \bibinfo {author} {\bibfnamefont {S.-C.}\ \bibnamefont {Zhang}},\
  }\href {\doibase 10.1103/RevModPhys.83.1057} {\bibfield  {journal} {\bibinfo
  {journal} {Rev. Mod. Phys.}\ }\textbf {\bibinfo {volume} {83}},\ \bibinfo
  {pages} {1057} (\bibinfo {year} {2011})}\BibitemShut {NoStop}%
\bibitem [{\citenamefont {Haldane}(1988)}]{PhysRevLett.61.2015}%
  \BibitemOpen
  \bibfield  {author} {\bibinfo {author} {\bibfnamefont {F.~D.~M.}\
  \bibnamefont {Haldane}},\ }\href {\doibase 10.1103/PhysRevLett.61.2015}
  {\bibfield  {journal} {\bibinfo  {journal} {Phys. Rev. Lett.}\ }\textbf
  {\bibinfo {volume} {61}},\ \bibinfo {pages} {2015} (\bibinfo {year}
  {1988})}\BibitemShut {NoStop}%
\bibitem [{\citenamefont {Kane}\ and\ \citenamefont
  {Mele}(2005)}]{PhysRevLett.95.226801}%
  \BibitemOpen
  \bibfield  {author} {\bibinfo {author} {\bibfnamefont {C.~L.}\ \bibnamefont
  {Kane}}\ and\ \bibinfo {author} {\bibfnamefont {E.~J.}\ \bibnamefont
  {Mele}},\ }\href {\doibase 10.1103/PhysRevLett.95.226801} {\bibfield
  {journal} {\bibinfo  {journal} {Phys. Rev. Lett.}\ }\textbf {\bibinfo
  {volume} {95}},\ \bibinfo {pages} {226801} (\bibinfo {year}
  {2005})}\BibitemShut {NoStop}%
\bibitem [{\citenamefont {Qi}\ \emph {et~al.}(2008)\citenamefont {Qi},
  \citenamefont {Hughes},\ and\ \citenamefont {Zhang}}]{PhysRevB.78.195424}%
  \BibitemOpen
  \bibfield  {author} {\bibinfo {author} {\bibfnamefont {X.-L.}\ \bibnamefont
  {Qi}}, \bibinfo {author} {\bibfnamefont {T.~L.}\ \bibnamefont {Hughes}}, \
  and\ \bibinfo {author} {\bibfnamefont {S.-C.}\ \bibnamefont {Zhang}},\ }\href
  {\doibase 10.1103/PhysRevB.78.195424} {\bibfield  {journal} {\bibinfo
  {journal} {Phys. Rev. B}\ }\textbf {\bibinfo {volume} {78}},\ \bibinfo
  {pages} {195424} (\bibinfo {year} {2008})}\BibitemShut {NoStop}%
\bibitem [{\citenamefont {Essin}\ \emph {et~al.}(2009)\citenamefont {Essin},
  \citenamefont {Moore},\ and\ \citenamefont
  {Vanderbilt}}]{PhysRevLett.102.146805}%
  \BibitemOpen
  \bibfield  {author} {\bibinfo {author} {\bibfnamefont {A.~M.}\ \bibnamefont
  {Essin}}, \bibinfo {author} {\bibfnamefont {J.~E.}\ \bibnamefont {Moore}}, \
  and\ \bibinfo {author} {\bibfnamefont {D.}~\bibnamefont {Vanderbilt}},\
  }\href {\doibase 10.1103/PhysRevLett.102.146805} {\bibfield  {journal}
  {\bibinfo  {journal} {Phys. Rev. Lett.}\ }\textbf {\bibinfo {volume} {102}},\
  \bibinfo {pages} {146805} (\bibinfo {year} {2009})}\BibitemShut {NoStop}%
\bibitem [{\citenamefont {Hatsugai}(1993{\natexlab{a}})}]{PhysRevLett.71.3697}%
  \BibitemOpen
  \bibfield  {author} {\bibinfo {author} {\bibfnamefont {Y.}~\bibnamefont
  {Hatsugai}},\ }\href {\doibase 10.1103/PhysRevLett.71.3697} {\bibfield
  {journal} {\bibinfo  {journal} {Phys. Rev. Lett.}\ }\textbf {\bibinfo
  {volume} {71}},\ \bibinfo {pages} {3697} (\bibinfo {year}
  {1993}{\natexlab{a}})}\BibitemShut {NoStop}%
\bibitem [{\citenamefont {Hatsugai}(1993{\natexlab{b}})}]{PhysRevB.48.11851}%
  \BibitemOpen
  \bibfield  {author} {\bibinfo {author} {\bibfnamefont {Y.}~\bibnamefont
  {Hatsugai}},\ }\href {\doibase 10.1103/PhysRevB.48.11851} {\bibfield
  {journal} {\bibinfo  {journal} {Phys. Rev. B}\ }\textbf {\bibinfo {volume}
  {48}},\ \bibinfo {pages} {11851} (\bibinfo {year}
  {1993}{\natexlab{b}})}\BibitemShut {NoStop}%
\bibitem [{\citenamefont {Hayashi}(2018)}]{Hayashi2018}%
  \BibitemOpen
  \bibfield  {author} {\bibinfo {author} {\bibfnamefont {S.}~\bibnamefont
  {Hayashi}},\ }\href {\doibase 10.1007/s00220-018-3229-2} {\bibfield
  {journal} {\bibinfo  {journal} {Commun. Math. Phys.}\ }\textbf {\bibinfo
  {volume} {364}},\ \bibinfo {pages} {343} (\bibinfo {year}
  {2018})}\BibitemShut {NoStop}%
\bibitem [{\citenamefont {Benalcazar}\ \emph
  {et~al.}(2017{\natexlab{a}})\citenamefont {Benalcazar}, \citenamefont
  {Bernevig},\ and\ \citenamefont {Hughes}}]{Benalcazar61}%
  \BibitemOpen
  \bibfield  {author} {\bibinfo {author} {\bibfnamefont {W.~A.}\ \bibnamefont
  {Benalcazar}}, \bibinfo {author} {\bibfnamefont {B.~A.}\ \bibnamefont
  {Bernevig}}, \ and\ \bibinfo {author} {\bibfnamefont {T.~L.}\ \bibnamefont
  {Hughes}},\ }\href {\doibase 10.1126/science.aah6442} {\bibfield  {journal}
  {\bibinfo  {journal} {Science}\ }\textbf {\bibinfo {volume} {357}},\ \bibinfo
  {pages} {61} (\bibinfo {year} {2017}{\natexlab{a}})}\BibitemShut {NoStop}%
\bibitem [{\citenamefont {Benalcazar}\ \emph
  {et~al.}(2017{\natexlab{b}})\citenamefont {Benalcazar}, \citenamefont
  {Bernevig},\ and\ \citenamefont {Hughes}}]{PhysRevB.96.245115}%
  \BibitemOpen
  \bibfield  {author} {\bibinfo {author} {\bibfnamefont {W.~A.}\ \bibnamefont
  {Benalcazar}}, \bibinfo {author} {\bibfnamefont {B.~A.}\ \bibnamefont
  {Bernevig}}, \ and\ \bibinfo {author} {\bibfnamefont {T.~L.}\ \bibnamefont
  {Hughes}},\ }\href {\doibase 10.1103/PhysRevB.96.245115} {\bibfield
  {journal} {\bibinfo  {journal} {Phys. Rev. B}\ }\textbf {\bibinfo {volume}
  {96}},\ \bibinfo {pages} {245115} (\bibinfo {year}
  {2017}{\natexlab{b}})}\BibitemShut {NoStop}%
\bibitem [{\citenamefont {Ezawa}(2018{\natexlab{a}})}]{PhysRevLett.120.026801}%
  \BibitemOpen
  \bibfield  {author} {\bibinfo {author} {\bibfnamefont {M.}~\bibnamefont
  {Ezawa}},\ }\href {\doibase 10.1103/PhysRevLett.120.026801} {\bibfield
  {journal} {\bibinfo  {journal} {Phys. Rev. Lett.}\ }\textbf {\bibinfo
  {volume} {120}},\ \bibinfo {pages} {026801} (\bibinfo {year}
  {2018}{\natexlab{a}})}\BibitemShut {NoStop}%
\bibitem [{\citenamefont {Schindler}\ \emph
  {et~al.}(2018{\natexlab{a}})\citenamefont {Schindler}, \citenamefont {Cook},
  \citenamefont {Vergniory}, \citenamefont {Wang}, \citenamefont {Parkin},
  \citenamefont {Bernevig},\ and\ \citenamefont {Neupert}}]{Schindlereaat0346}%
  \BibitemOpen
  \bibfield  {author} {\bibinfo {author} {\bibfnamefont {F.}~\bibnamefont
  {Schindler}}, \bibinfo {author} {\bibfnamefont {A.~M.}\ \bibnamefont {Cook}},
  \bibinfo {author} {\bibfnamefont {M.~G.}\ \bibnamefont {Vergniory}}, \bibinfo
  {author} {\bibfnamefont {Z.}~\bibnamefont {Wang}}, \bibinfo {author}
  {\bibfnamefont {S.~S.~P.}\ \bibnamefont {Parkin}}, \bibinfo {author}
  {\bibfnamefont {B.~A.}\ \bibnamefont {Bernevig}}, \ and\ \bibinfo {author}
  {\bibfnamefont {T.}~\bibnamefont {Neupert}},\ }\href {\doibase
  10.1126/sciadv.aat0346} {\bibfield  {journal} {\bibinfo  {journal} {Sci.
  Adv.}\ }\textbf {\bibinfo {volume} {4}},\ \bibinfo {pages} {eaat0346}
  (\bibinfo {year} {2018}{\natexlab{a}})}\BibitemShut {NoStop}%
\bibitem [{\citenamefont {Kunst}\ \emph {et~al.}(2018)\citenamefont {Kunst},
  \citenamefont {van Miert},\ and\ \citenamefont
  {Bergholtz}}]{PhysRevB.97.241405}%
  \BibitemOpen
  \bibfield  {author} {\bibinfo {author} {\bibfnamefont {F.~K.}\ \bibnamefont
  {Kunst}}, \bibinfo {author} {\bibfnamefont {G.}~\bibnamefont {van Miert}}, \
  and\ \bibinfo {author} {\bibfnamefont {E.~J.}\ \bibnamefont {Bergholtz}},\
  }\href {\doibase 10.1103/PhysRevB.97.241405} {\bibfield  {journal} {\bibinfo
  {journal} {Phys. Rev. B}\ }\textbf {\bibinfo {volume} {97}},\ \bibinfo
  {pages} {241405(R)} (\bibinfo {year} {2018})}\BibitemShut {NoStop}%
\bibitem [{\citenamefont {Xu}\ \emph {et~al.}()\citenamefont {Xu},
  \citenamefont {Xue},\ and\ \citenamefont {Wan}}]{2017arXiv171109202X}%
  \BibitemOpen
  \bibfield  {author} {\bibinfo {author} {\bibfnamefont {Y.}~\bibnamefont
  {Xu}}, \bibinfo {author} {\bibfnamefont {R.}~\bibnamefont {Xue}}, \ and\
  \bibinfo {author} {\bibfnamefont {S.}~\bibnamefont {Wan}},\ }\href
  {https://arxiv.org/abs/1711.09202} {\ }\Eprint
  {http://arxiv.org/abs/1711.09202} {arXiv:1711.09202} \BibitemShut {NoStop}%
\bibitem [{\citenamefont {C\ifmmode \u{a}\else \u{a}\fi{}lug\ifmmode~\u{a}\else
  \u{a}\fi{}ru}\ \emph {et~al.}(2019)\citenamefont {C\ifmmode \u{a}\else
  \u{a}\fi{}lug\ifmmode~\u{a}\else \u{a}\fi{}ru}, \citenamefont {Juri\ifmmode
  \check{c}\else \v{c}\fi{}i\ifmmode~\acute{c}\else \'{c}\fi{}},\ and\
  \citenamefont {Roy}}]{PhysRevB.99.041301}%
  \BibitemOpen
  \bibfield  {author} {\bibinfo {author} {\bibfnamefont {D.}~\bibnamefont
  {C\ifmmode \u{a}\else \u{a}\fi{}lug\ifmmode~\u{a}\else \u{a}\fi{}ru}},
  \bibinfo {author} {\bibfnamefont {V.}~\bibnamefont {Juri\ifmmode
  \check{c}\else \v{c}\fi{}i\ifmmode~\acute{c}\else \'{c}\fi{}}}, \ and\
  \bibinfo {author} {\bibfnamefont {B.}~\bibnamefont {Roy}},\ }\href {\doibase
  10.1103/PhysRevB.99.041301} {\bibfield  {journal} {\bibinfo  {journal} {Phys.
  Rev. B}\ }\textbf {\bibinfo {volume} {99}},\ \bibinfo {pages} {041301(R)}
  (\bibinfo {year} {2019})}\BibitemShut {NoStop}%
\bibitem [{\citenamefont {Kang}\ \emph {et~al.}(2019)\citenamefont {Kang},
  \citenamefont {Shiozaki},\ and\ \citenamefont {Cho}}]{PhysRevB.100.245134}%
  \BibitemOpen
  \bibfield  {author} {\bibinfo {author} {\bibfnamefont {B.}~\bibnamefont
  {Kang}}, \bibinfo {author} {\bibfnamefont {K.}~\bibnamefont {Shiozaki}}, \
  and\ \bibinfo {author} {\bibfnamefont {G.~Y.}\ \bibnamefont {Cho}},\ }\href
  {\doibase 10.1103/PhysRevB.100.245134} {\bibfield  {journal} {\bibinfo
  {journal} {Phys. Rev. B}\ }\textbf {\bibinfo {volume} {100}},\ \bibinfo
  {pages} {245134} (\bibinfo {year} {2019})}\BibitemShut {NoStop}%
\bibitem [{\citenamefont {Fukui}\ and\ \citenamefont
  {Hatsugai}(2018)}]{PhysRevB.98.035147}%
  \BibitemOpen
  \bibfield  {author} {\bibinfo {author} {\bibfnamefont {T.}~\bibnamefont
  {Fukui}}\ and\ \bibinfo {author} {\bibfnamefont {Y.}~\bibnamefont
  {Hatsugai}},\ }\href {\doibase 10.1103/PhysRevB.98.035147} {\bibfield
  {journal} {\bibinfo  {journal} {Phys. Rev. B}\ }\textbf {\bibinfo {volume}
  {98}},\ \bibinfo {pages} {035147} (\bibinfo {year} {2018})}\BibitemShut
  {NoStop}%
\bibitem [{\citenamefont {Hatsugai}\ and\ \citenamefont
  {Maruyama}(2011)}]{Hatsugai_2011}%
  \BibitemOpen
  \bibfield  {author} {\bibinfo {author} {\bibfnamefont {Y.}~\bibnamefont
  {Hatsugai}}\ and\ \bibinfo {author} {\bibfnamefont {I.}~\bibnamefont
  {Maruyama}},\ }\href {\doibase 10.1209/0295-5075/95/20003} {\bibfield
  {journal} {\bibinfo  {journal} {Europhys. Lett.}\ }\textbf {\bibinfo {volume}
  {95}},\ \bibinfo {pages} {20003} (\bibinfo {year} {2011})}\BibitemShut
  {NoStop}%
\bibitem [{\citenamefont {Kudo}\ \emph {et~al.}(2019)\citenamefont {Kudo},
  \citenamefont {Yoshida},\ and\ \citenamefont
  {Hatsugai}}]{PhysRevLett.123.196402}%
  \BibitemOpen
  \bibfield  {author} {\bibinfo {author} {\bibfnamefont {K.}~\bibnamefont
  {Kudo}}, \bibinfo {author} {\bibfnamefont {T.}~\bibnamefont {Yoshida}}, \
  and\ \bibinfo {author} {\bibfnamefont {Y.}~\bibnamefont {Hatsugai}},\ }\href
  {\doibase 10.1103/PhysRevLett.123.196402} {\bibfield  {journal} {\bibinfo
  {journal} {Phys. Rev. Lett.}\ }\textbf {\bibinfo {volume} {123}},\ \bibinfo
  {pages} {196402} (\bibinfo {year} {2019})}\BibitemShut {NoStop}%
\bibitem [{\citenamefont {Araki}\ \emph {et~al.}(2020)\citenamefont {Araki},
  \citenamefont {Mizoguchi},\ and\ \citenamefont
  {Hatsugai}}]{PhysRevResearch.2.012009}%
  \BibitemOpen
  \bibfield  {author} {\bibinfo {author} {\bibfnamefont {H.}~\bibnamefont
  {Araki}}, \bibinfo {author} {\bibfnamefont {T.}~\bibnamefont {Mizoguchi}}, \
  and\ \bibinfo {author} {\bibfnamefont {Y.}~\bibnamefont {Hatsugai}},\ }\href
  {\doibase 10.1103/PhysRevResearch.2.012009} {\bibfield  {journal} {\bibinfo
  {journal} {Phys. Rev. Research}\ }\textbf {\bibinfo {volume} {2}},\ \bibinfo
  {pages} {012009} (\bibinfo {year} {2020})}\BibitemShut {NoStop}%
\bibitem [{\citenamefont {Ezawa}(2018{\natexlab{b}})}]{PhysRevB.98.045125}%
  \BibitemOpen
  \bibfield  {author} {\bibinfo {author} {\bibfnamefont {M.}~\bibnamefont
  {Ezawa}},\ }\href {\doibase 10.1103/PhysRevB.98.045125} {\bibfield  {journal}
  {\bibinfo  {journal} {Phys. Rev. B}\ }\textbf {\bibinfo {volume} {98}},\
  \bibinfo {pages} {045125} (\bibinfo {year} {2018}{\natexlab{b}})}\BibitemShut
  {NoStop}%
\bibitem [{\citenamefont {Schindler}\ \emph
  {et~al.}(2018{\natexlab{b}})\citenamefont {Schindler}, \citenamefont {Wang},
  \citenamefont {Vergniory}, \citenamefont {Cook}, \citenamefont {Murani},
  \citenamefont {Sengupta}, \citenamefont {Kasumov}, \citenamefont {Deblock},
  \citenamefont {Jeon}, \citenamefont {Drozdov}, \citenamefont {Bouchiat},
  \citenamefont {Gu\'{e}ron}, \citenamefont {Yazdani}, \citenamefont
  {Bernevig},\ and\ \citenamefont {Neupert}}]{schindler2018higher}%
  \BibitemOpen
  \bibfield  {author} {\bibinfo {author} {\bibfnamefont {F.}~\bibnamefont
  {Schindler}}, \bibinfo {author} {\bibfnamefont {Z.}~\bibnamefont {Wang}},
  \bibinfo {author} {\bibfnamefont {M.~G.}\ \bibnamefont {Vergniory}}, \bibinfo
  {author} {\bibfnamefont {A.~M.}\ \bibnamefont {Cook}}, \bibinfo {author}
  {\bibfnamefont {A.}~\bibnamefont {Murani}}, \bibinfo {author} {\bibfnamefont
  {S.}~\bibnamefont {Sengupta}}, \bibinfo {author} {\bibfnamefont {A.~Y.}\
  \bibnamefont {Kasumov}}, \bibinfo {author} {\bibfnamefont {R.}~\bibnamefont
  {Deblock}}, \bibinfo {author} {\bibfnamefont {S.}~\bibnamefont {Jeon}},
  \bibinfo {author} {\bibfnamefont {I.}~\bibnamefont {Drozdov}}, \bibinfo
  {author} {\bibfnamefont {H.}~\bibnamefont {Bouchiat}}, \bibinfo {author}
  {\bibfnamefont {S.}~\bibnamefont {Gu\'{e}ron}}, \bibinfo {author}
  {\bibfnamefont {A.}~\bibnamefont {Yazdani}}, \bibinfo {author} {\bibfnamefont
  {B.~A.}\ \bibnamefont {Bernevig}}, \ and\ \bibinfo {author} {\bibfnamefont
  {T.}~\bibnamefont {Neupert}},\ }\href {\doibase 10.1038/s41567-018-0224-7}
  {\bibfield  {journal} {\bibinfo  {journal} {Nat. Phys.}\ }\textbf {\bibinfo
  {volume} {14}},\ \bibinfo {pages} {918} (\bibinfo {year}
  {2018}{\natexlab{b}})}\BibitemShut {NoStop}%
\bibitem [{\citenamefont {Sheng}\ \emph {et~al.}(2019)\citenamefont {Sheng},
  \citenamefont {Chen}, \citenamefont {Liu}, \citenamefont {Chen},
  \citenamefont {Yu}, \citenamefont {Zhao},\ and\ \citenamefont
  {Yang}}]{PhysRevLett123256402}%
  \BibitemOpen
  \bibfield  {author} {\bibinfo {author} {\bibfnamefont {X.-L.}\ \bibnamefont
  {Sheng}}, \bibinfo {author} {\bibfnamefont {C.}~\bibnamefont {Chen}},
  \bibinfo {author} {\bibfnamefont {H.}~\bibnamefont {Liu}}, \bibinfo {author}
  {\bibfnamefont {Z.}~\bibnamefont {Chen}}, \bibinfo {author} {\bibfnamefont
  {Z.-M.}\ \bibnamefont {Yu}}, \bibinfo {author} {\bibfnamefont {Y.~X.}\
  \bibnamefont {Zhao}}, \ and\ \bibinfo {author} {\bibfnamefont {S.~A.}\
  \bibnamefont {Yang}},\ }\href {\doibase 10.1103/PhysRevLett.123.256402}
  {\bibfield  {journal} {\bibinfo  {journal} {Phys. Rev. Lett.}\ }\textbf
  {\bibinfo {volume} {123}},\ \bibinfo {pages} {256402} (\bibinfo {year}
  {2019})}\BibitemShut {NoStop}%
\bibitem [{\citenamefont {Lee}\ \emph {et~al.}(2020)\citenamefont {Lee},
  \citenamefont {Kim}, \citenamefont {Ahn},\ and\ \citenamefont
  {Yang}}]{2019arXiv190411452L}%
  \BibitemOpen
  \bibfield  {author} {\bibinfo {author} {\bibfnamefont {E.}~\bibnamefont
  {Lee}}, \bibinfo {author} {\bibfnamefont {R.}~\bibnamefont {Kim}}, \bibinfo
  {author} {\bibfnamefont {J.}~\bibnamefont {Ahn}}, \ and\ \bibinfo {author}
  {\bibfnamefont {B.-J.}\ \bibnamefont {Yang}},\ }\href {\doibase
  10.1038/s41535-019-0206-8} {\bibfield  {journal} {\bibinfo  {journal} {npj
  Quantum Mater.}\ }\textbf {\bibinfo {volume} {5}},\ \bibinfo {pages} {1}
  (\bibinfo {year} {2020})}\BibitemShut {NoStop}%
\bibitem [{\citenamefont {{Lee}}\ \emph {et~al.}()\citenamefont {{Lee}},
  \citenamefont {{Geng}}, \citenamefont {{Park}}, \citenamefont {{Oshikawa}},
  \citenamefont {{Lee}}, \citenamefont {{Yeom}},\ and\ \citenamefont
  {{Cho}}}]{2019arXiv190700012L}%
  \BibitemOpen
  \bibfield  {author} {\bibinfo {author} {\bibfnamefont {J.~M.}\ \bibnamefont
  {{Lee}}}, \bibinfo {author} {\bibfnamefont {C.}~\bibnamefont {{Geng}}},
  \bibinfo {author} {\bibfnamefont {J.~W.}\ \bibnamefont {{Park}}}, \bibinfo
  {author} {\bibfnamefont {M.}~\bibnamefont {{Oshikawa}}}, \bibinfo {author}
  {\bibfnamefont {S.-S.}\ \bibnamefont {{Lee}}}, \bibinfo {author}
  {\bibfnamefont {H.~W.}\ \bibnamefont {{Yeom}}}, \ and\ \bibinfo {author}
  {\bibfnamefont {G.~Y.}\ \bibnamefont {{Cho}}},\ }\href
  {https://arxiv.org/abs/1907.00012} {\ }\Eprint
  {http://arxiv.org/abs/1907.00012} {arXiv:1907.00012} \BibitemShut {NoStop}%
\bibitem [{\citenamefont {Mizoguchi}\ \emph
  {et~al.}(2019{\natexlab{a}})\citenamefont {Mizoguchi}, \citenamefont
  {Maruyama}, \citenamefont {Okada},\ and\ \citenamefont
  {Hatsugai}}]{PhysRevMaterials.3.114201}%
  \BibitemOpen
  \bibfield  {author} {\bibinfo {author} {\bibfnamefont {T.}~\bibnamefont
  {Mizoguchi}}, \bibinfo {author} {\bibfnamefont {M.}~\bibnamefont {Maruyama}},
  \bibinfo {author} {\bibfnamefont {S.}~\bibnamefont {Okada}}, \ and\ \bibinfo
  {author} {\bibfnamefont {Y.}~\bibnamefont {Hatsugai}},\ }\href {\doibase
  10.1103/PhysRevMaterials.3.114201} {\bibfield  {journal} {\bibinfo  {journal}
  {Phys. Rev. Materials}\ }\textbf {\bibinfo {volume} {3}},\ \bibinfo {pages}
  {114201} (\bibinfo {year} {2019}{\natexlab{a}})}\BibitemShut {NoStop}%
\bibitem [{\citenamefont {Serra-Garcia}\ \emph {et~al.}(2018)\citenamefont
  {Serra-Garcia}, \citenamefont {Peri}, \citenamefont {S{\"u}sstrunk},
  \citenamefont {Bilal}, \citenamefont {Larsen}, \citenamefont {Villanueva},\
  and\ \citenamefont {Huber}}]{serra2018observation}%
  \BibitemOpen
  \bibfield  {author} {\bibinfo {author} {\bibfnamefont {M.}~\bibnamefont
  {Serra-Garcia}}, \bibinfo {author} {\bibfnamefont {V.}~\bibnamefont {Peri}},
  \bibinfo {author} {\bibfnamefont {R.}~\bibnamefont {S{\"u}sstrunk}}, \bibinfo
  {author} {\bibfnamefont {O.~R.}\ \bibnamefont {Bilal}}, \bibinfo {author}
  {\bibfnamefont {T.}~\bibnamefont {Larsen}}, \bibinfo {author} {\bibfnamefont
  {L.~G.}\ \bibnamefont {Villanueva}}, \ and\ \bibinfo {author} {\bibfnamefont
  {S.~D.}\ \bibnamefont {Huber}},\ }\href
  {https://www.nature.com/articles/nature25156} {\bibfield  {journal} {\bibinfo
   {journal} {Nature}\ }\textbf {\bibinfo {volume} {555}},\ \bibinfo {pages}
  {342} (\bibinfo {year} {2018})}\BibitemShut {NoStop}%
\bibitem [{\citenamefont {Attig}\ \emph {et~al.}(2019)\citenamefont {Attig},
  \citenamefont {Roychowdhury}, \citenamefont {Lawler},\ and\ \citenamefont
  {Trebst}}]{PhysRevResearch.1.032047}%
  \BibitemOpen
  \bibfield  {author} {\bibinfo {author} {\bibfnamefont {J.}~\bibnamefont
  {Attig}}, \bibinfo {author} {\bibfnamefont {K.}~\bibnamefont {Roychowdhury}},
  \bibinfo {author} {\bibfnamefont {M.~J.}\ \bibnamefont {Lawler}}, \ and\
  \bibinfo {author} {\bibfnamefont {S.}~\bibnamefont {Trebst}},\ }\href
  {\doibase 10.1103/PhysRevResearch.1.032047} {\bibfield  {journal} {\bibinfo
  {journal} {Phys. Rev. Research}\ }\textbf {\bibinfo {volume} {1}},\ \bibinfo
  {pages} {032047} (\bibinfo {year} {2019})}\BibitemShut {NoStop}%
\bibitem [{\citenamefont {El~Hassan}\ \emph {et~al.}(2019)\citenamefont
  {El~Hassan}, \citenamefont {Kunst}, \citenamefont {Moritz}, \citenamefont
  {Andler}, \citenamefont {Bergholtz},\ and\ \citenamefont
  {Bourennane}}]{ElHassan2019}%
  \BibitemOpen
  \bibfield  {author} {\bibinfo {author} {\bibfnamefont {A.}~\bibnamefont
  {El~Hassan}}, \bibinfo {author} {\bibfnamefont {F.~K.}\ \bibnamefont
  {Kunst}}, \bibinfo {author} {\bibfnamefont {A.}~\bibnamefont {Moritz}},
  \bibinfo {author} {\bibfnamefont {G.}~\bibnamefont {Andler}}, \bibinfo
  {author} {\bibfnamefont {E.~J.}\ \bibnamefont {Bergholtz}}, \ and\ \bibinfo
  {author} {\bibfnamefont {M.}~\bibnamefont {Bourennane}},\ }\href {\doibase
  10.1038/s41566-019-0519-y} {\bibfield  {journal} {\bibinfo  {journal} {Nature
  Photonics}\ }\textbf {\bibinfo {volume} {13}},\ \bibinfo {pages} {697}
  (\bibinfo {year} {2019})}\BibitemShut {NoStop}%
\bibitem [{\citenamefont {Ota}\ \emph {et~al.}(2019)\citenamefont {Ota},
  \citenamefont {Liu}, \citenamefont {Katsumi}, \citenamefont {Watanabe},
  \citenamefont {Wakabayashi}, \citenamefont {Arakawa},\ and\ \citenamefont
  {Iwamoto}}]{Ota:19}%
  \BibitemOpen
  \bibfield  {author} {\bibinfo {author} {\bibfnamefont {Y.}~\bibnamefont
  {Ota}}, \bibinfo {author} {\bibfnamefont {F.}~\bibnamefont {Liu}}, \bibinfo
  {author} {\bibfnamefont {R.}~\bibnamefont {Katsumi}}, \bibinfo {author}
  {\bibfnamefont {K.}~\bibnamefont {Watanabe}}, \bibinfo {author}
  {\bibfnamefont {K.}~\bibnamefont {Wakabayashi}}, \bibinfo {author}
  {\bibfnamefont {Y.}~\bibnamefont {Arakawa}}, \ and\ \bibinfo {author}
  {\bibfnamefont {S.}~\bibnamefont {Iwamoto}},\ }\href {\doibase
  10.1364/OPTICA.6.000786} {\bibfield  {journal} {\bibinfo  {journal} {Optica}\
  }\textbf {\bibinfo {volume} {6}},\ \bibinfo {pages} {786} (\bibinfo {year}
  {2019})}\BibitemShut {NoStop}%
\bibitem [{\citenamefont {Ni}\ \emph {et~al.}(2017)\citenamefont {Ni},
  \citenamefont {Gorlach}, \citenamefont {Alu},\ and\ \citenamefont
  {Khanikaev}}]{Ni_2017}%
  \BibitemOpen
  \bibfield  {author} {\bibinfo {author} {\bibfnamefont {X.}~\bibnamefont
  {Ni}}, \bibinfo {author} {\bibfnamefont {M.~A.}\ \bibnamefont {Gorlach}},
  \bibinfo {author} {\bibfnamefont {A.}~\bibnamefont {Alu}}, \ and\ \bibinfo
  {author} {\bibfnamefont {A.~B.}\ \bibnamefont {Khanikaev}},\ }\href {\doibase
  10.1088/1367-2630/aa6996} {\bibfield  {journal} {\bibinfo  {journal} {N. J.
  Phys.}\ }\textbf {\bibinfo {volume} {19}},\ \bibinfo {pages} {055002}
  (\bibinfo {year} {2017})}\BibitemShut {NoStop}%
\bibitem [{\citenamefont {Xue}\ \emph {et~al.}(2019)\citenamefont {Xue},
  \citenamefont {Yang}, \citenamefont {Gao}, \citenamefont {Chong},\ and\
  \citenamefont {Zhang}}]{xue2019acoustic}%
  \BibitemOpen
  \bibfield  {author} {\bibinfo {author} {\bibfnamefont {H.}~\bibnamefont
  {Xue}}, \bibinfo {author} {\bibfnamefont {Y.}~\bibnamefont {Yang}}, \bibinfo
  {author} {\bibfnamefont {F.}~\bibnamefont {Gao}}, \bibinfo {author}
  {\bibfnamefont {Y.}~\bibnamefont {Chong}}, \ and\ \bibinfo {author}
  {\bibfnamefont {B.}~\bibnamefont {Zhang}},\ }\href {\doibase
  10.1038/s41563-018-0251-x} {\bibfield  {journal} {\bibinfo  {journal} {Nat.
  Mater.}\ }\textbf {\bibinfo {volume} {18}},\ \bibinfo {pages} {108} (\bibinfo
  {year} {2019})}\BibitemShut {NoStop}%
\bibitem [{\citenamefont {Ni}\ \emph {et~al.}(2019)\citenamefont {Ni},
  \citenamefont {Weiner}, \citenamefont {Al{\`u}},\ and\ \citenamefont
  {Khanikaev}}]{ni2019observation}%
  \BibitemOpen
  \bibfield  {author} {\bibinfo {author} {\bibfnamefont {X.}~\bibnamefont
  {Ni}}, \bibinfo {author} {\bibfnamefont {M.}~\bibnamefont {Weiner}}, \bibinfo
  {author} {\bibfnamefont {A.}~\bibnamefont {Al{\`u}}}, \ and\ \bibinfo
  {author} {\bibfnamefont {A.~B.}\ \bibnamefont {Khanikaev}},\ }\href {\doibase
  10.1038/s41563-018-0252-9} {\bibfield  {journal} {\bibinfo  {journal} {Nature
  materials}\ }\textbf {\bibinfo {volume} {18}},\ \bibinfo {pages} {113}
  (\bibinfo {year} {2019})}\BibitemShut {NoStop}%
\bibitem [{\citenamefont {Imhof}\ \emph {et~al.}(2018)\citenamefont {Imhof},
  \citenamefont {Berger}, \citenamefont {Bayer}, \citenamefont {Brehm},
  \citenamefont {Molenkamp}, \citenamefont {Kiessling}, \citenamefont
  {Schindler}, \citenamefont {Lee}, \citenamefont {Greiter}, \citenamefont
  {Neupert},\ and\ \citenamefont {Thomale}}]{imhof2018topolectrical}%
  \BibitemOpen
  \bibfield  {author} {\bibinfo {author} {\bibfnamefont {S.}~\bibnamefont
  {Imhof}}, \bibinfo {author} {\bibfnamefont {C.}~\bibnamefont {Berger}},
  \bibinfo {author} {\bibfnamefont {F.}~\bibnamefont {Bayer}}, \bibinfo
  {author} {\bibfnamefont {J.}~\bibnamefont {Brehm}}, \bibinfo {author}
  {\bibfnamefont {L.~W.}\ \bibnamefont {Molenkamp}}, \bibinfo {author}
  {\bibfnamefont {T.}~\bibnamefont {Kiessling}}, \bibinfo {author}
  {\bibfnamefont {F.}~\bibnamefont {Schindler}}, \bibinfo {author}
  {\bibfnamefont {C.~H.}\ \bibnamefont {Lee}}, \bibinfo {author} {\bibfnamefont
  {M.}~\bibnamefont {Greiter}}, \bibinfo {author} {\bibfnamefont
  {T.}~\bibnamefont {Neupert}}, \ and\ \bibinfo {author} {\bibfnamefont
  {R.}~\bibnamefont {Thomale}},\ }\href {\doibase 10.1038/s41567-018-0246-1}
  {\bibfield  {journal} {\bibinfo  {journal} {Nat. Phys.}\ }\textbf {\bibinfo
  {volume} {14}},\ \bibinfo {pages} {925} (\bibinfo {year} {2018})}\BibitemShut
  {NoStop}%
\bibitem [{\citenamefont {Ezawa}(2018{\natexlab{c}})}]{PhysRevB.98.201402}%
  \BibitemOpen
  \bibfield  {author} {\bibinfo {author} {\bibfnamefont {M.}~\bibnamefont
  {Ezawa}},\ }\href {\doibase 10.1103/PhysRevB.98.201402} {\bibfield  {journal}
  {\bibinfo  {journal} {Phys. Rev. B}\ }\textbf {\bibinfo {volume} {98}},\
  \bibinfo {pages} {201402(R)} (\bibinfo {year}
  {2018}{\natexlab{c}})}\BibitemShut {NoStop}%
\bibitem [{\citenamefont {Kempkes}\ \emph {et~al.}(2019)\citenamefont
  {Kempkes}, \citenamefont {Slot}, \citenamefont {van~den Broeke},
  \citenamefont {Capiod}, \citenamefont {Benalcazar}, \citenamefont
  {Vanmaekelbergh}, \citenamefont {Bercioux}, \citenamefont {Swart},\ and\
  \citenamefont {Morais~Smith}}]{Kempkes2019observation}%
  \BibitemOpen
  \bibfield  {author} {\bibinfo {author} {\bibfnamefont {S.~N.}\ \bibnamefont
  {Kempkes}}, \bibinfo {author} {\bibfnamefont {M.~R.}\ \bibnamefont {Slot}},
  \bibinfo {author} {\bibfnamefont {J.~J.}\ \bibnamefont {van~den Broeke}},
  \bibinfo {author} {\bibfnamefont {P.}~\bibnamefont {Capiod}}, \bibinfo
  {author} {\bibfnamefont {W.~A.}\ \bibnamefont {Benalcazar}}, \bibinfo
  {author} {\bibfnamefont {D.}~\bibnamefont {Vanmaekelbergh}}, \bibinfo
  {author} {\bibfnamefont {D.}~\bibnamefont {Bercioux}}, \bibinfo {author}
  {\bibfnamefont {I.}~\bibnamefont {Swart}}, \ and\ \bibinfo {author}
  {\bibfnamefont {C.}~\bibnamefont {Morais~Smith}},\ }\href {\doibase
  10.1038/s41563-019-0483-4} {\bibfield  {journal} {\bibinfo  {journal} {Nature
  Materials}\ }\textbf {\bibinfo {volume} {18}},\ \bibinfo {pages} {1292}
  (\bibinfo {year} {2019})}\BibitemShut {NoStop}%
\bibitem [{\citenamefont {Kane}\ and\ \citenamefont
  {Lubensky}(2014)}]{kane2014topological}%
  \BibitemOpen
  \bibfield  {author} {\bibinfo {author} {\bibfnamefont {C.}~\bibnamefont
  {Kane}}\ and\ \bibinfo {author} {\bibfnamefont {T.}~\bibnamefont
  {Lubensky}},\ }\href {\doibase 10.1038/nphys2835} {\bibfield  {journal}
  {\bibinfo  {journal} {Nat. Phys.}\ }\textbf {\bibinfo {volume} {10}},\
  \bibinfo {pages} {39} (\bibinfo {year} {2014})}\BibitemShut {NoStop}%
\bibitem [{\citenamefont {Wang}\ \emph {et~al.}(2015)\citenamefont {Wang},
  \citenamefont {Luan},\ and\ \citenamefont {Zhang}}]{Wang_2015}%
  \BibitemOpen
  \bibfield  {author} {\bibinfo {author} {\bibfnamefont {Y.-T.}\ \bibnamefont
  {Wang}}, \bibinfo {author} {\bibfnamefont {P.-G.}\ \bibnamefont {Luan}}, \
  and\ \bibinfo {author} {\bibfnamefont {S.}~\bibnamefont {Zhang}},\ }\href
  {\doibase 10.1088/1367-2630/17/7/073031} {\bibfield  {journal} {\bibinfo
  {journal} {N. J. Phys.}\ }\textbf {\bibinfo {volume} {17}},\ \bibinfo {pages}
  {073031} (\bibinfo {year} {2015})}\BibitemShut {NoStop}%
\bibitem [{\citenamefont {Kariyado}\ and\ \citenamefont
  {Hatsugai}(2015)}]{kariyado2015manipulation}%
  \BibitemOpen
  \bibfield  {author} {\bibinfo {author} {\bibfnamefont {T.}~\bibnamefont
  {Kariyado}}\ and\ \bibinfo {author} {\bibfnamefont {Y.}~\bibnamefont
  {Hatsugai}},\ }\href {\doibase 10.1038/srep18107 (2015).} {\bibfield
  {journal} {\bibinfo  {journal} {Sci. Rep.}\ }\textbf {\bibinfo {volume}
  {5}},\ \bibinfo {pages} {18107} (\bibinfo {year} {2015})}\BibitemShut
  {NoStop}%
\bibitem [{\citenamefont {Takahashi}\ \emph {et~al.}(2017)\citenamefont
  {Takahashi}, \citenamefont {Kariyado},\ and\ \citenamefont
  {Hatsugai}}]{takahashi2017edge}%
  \BibitemOpen
  \bibfield  {author} {\bibinfo {author} {\bibfnamefont {Y.}~\bibnamefont
  {Takahashi}}, \bibinfo {author} {\bibfnamefont {T.}~\bibnamefont {Kariyado}},
  \ and\ \bibinfo {author} {\bibfnamefont {Y.}~\bibnamefont {Hatsugai}},\
  }\href {https://iopscience.iop.org/article/10.1088/1367-2630/aa5edb}
  {\bibfield  {journal} {\bibinfo  {journal} {N. J. Phys.}\ }\textbf {\bibinfo
  {volume} {19}},\ \bibinfo {pages} {035003} (\bibinfo {year}
  {2017})}\BibitemShut {NoStop}%
\bibitem [{\citenamefont {Takahashi}\ \emph {et~al.}(2019)\citenamefont
  {Takahashi}, \citenamefont {Kariyado},\ and\ \citenamefont
  {Hatsugai}}]{PhysRevB.99.024102}%
  \BibitemOpen
  \bibfield  {author} {\bibinfo {author} {\bibfnamefont {Y.}~\bibnamefont
  {Takahashi}}, \bibinfo {author} {\bibfnamefont {T.}~\bibnamefont {Kariyado}},
  \ and\ \bibinfo {author} {\bibfnamefont {Y.}~\bibnamefont {Hatsugai}},\
  }\href {\doibase 10.1103/PhysRevB.99.024102} {\bibfield  {journal} {\bibinfo
  {journal} {Phys. Rev. B}\ }\textbf {\bibinfo {volume} {99}},\ \bibinfo
  {pages} {024102} (\bibinfo {year} {2019})}\BibitemShut {NoStop}%
\bibitem [{\citenamefont {Yoshida}\ and\ \citenamefont
  {Hatsugai}(2019)}]{PhysRevB.100.054109}%
  \BibitemOpen
  \bibfield  {author} {\bibinfo {author} {\bibfnamefont {T.}~\bibnamefont
  {Yoshida}}\ and\ \bibinfo {author} {\bibfnamefont {Y.}~\bibnamefont
  {Hatsugai}},\ }\href {\doibase 10.1103/PhysRevB.100.054109} {\bibfield
  {journal} {\bibinfo  {journal} {Phys. Rev. B}\ }\textbf {\bibinfo {volume}
  {100}},\ \bibinfo {pages} {054109} (\bibinfo {year} {2019})}\BibitemShut
  {NoStop}%
\bibitem [{foo({\natexlab{a}})}]{footnote_one}%
  \BibitemOpen
  \href@noop {} {}\bibinfo {note} {For $\eta_{\alpha}=0$, we consider the
  situation where $l_{\alpha} \rightarrow 0$ with keeping $R_{\alpha}$
  unity}\BibitemShut {NoStop}%
\bibitem [{\citenamefont {Hatsugai}(2006)}]{200615354}%
  \BibitemOpen
  \bibfield  {author} {\bibinfo {author} {\bibfnamefont {Y.}~\bibnamefont
  {Hatsugai}},\ }\href {\doibase 10.1143/jpsj.75.123601} {\bibfield  {journal}
  {\bibinfo  {journal} {J. Phys. Soc. Jpn.}\ }\textbf {\bibinfo {volume}
  {75}},\ \bibinfo {pages} {123601} (\bibinfo {year} {2006})}\BibitemShut
  {NoStop}%
\bibitem [{\citenamefont {Hatsugai}(2007)}]{Hatsugai_2007}%
  \BibitemOpen
  \bibfield  {author} {\bibinfo {author} {\bibfnamefont {Y.}~\bibnamefont
  {Hatsugai}},\ }\href {\doibase 10.1088/0953-8984/19/14/145209} {\bibfield
  {journal} {\bibinfo  {journal} {J. Phys. Condens. Matter}\ }\textbf {\bibinfo
  {volume} {19}},\ \bibinfo {pages} {145209} (\bibinfo {year}
  {2007})}\BibitemShut {NoStop}%
\bibitem [{\citenamefont {Hirano}\ \emph {et~al.}(2008)\citenamefont {Hirano},
  \citenamefont {Katsura},\ and\ \citenamefont
  {Hatsugai}}]{PhysRevB.77.094431}%
  \BibitemOpen
  \bibfield  {author} {\bibinfo {author} {\bibfnamefont {T.}~\bibnamefont
  {Hirano}}, \bibinfo {author} {\bibfnamefont {H.}~\bibnamefont {Katsura}}, \
  and\ \bibinfo {author} {\bibfnamefont {Y.}~\bibnamefont {Hatsugai}},\ }\href
  {\doibase 10.1103/PhysRevB.77.094431} {\bibfield  {journal} {\bibinfo
  {journal} {Phys. Rev. B}\ }\textbf {\bibinfo {volume} {77}},\ \bibinfo
  {pages} {094431} (\bibinfo {year} {2008})}\BibitemShut {NoStop}%
\bibitem [{\citenamefont {Hatsugai}(2010)}]{Hatsugai_2010}%
  \BibitemOpen
  \bibfield  {author} {\bibinfo {author} {\bibfnamefont {Y.}~\bibnamefont
  {Hatsugai}},\ }\href {\doibase 10.1088/1367-2630/12/6/065004} {\bibfield
  {journal} {\bibinfo  {journal} {N. J. Phys.}\ }\textbf {\bibinfo {volume}
  {12}},\ \bibinfo {pages} {065004} (\bibinfo {year} {2010})}\BibitemShut
  {NoStop}%
\bibitem [{\citenamefont {Motoyama}\ and\ \citenamefont
  {Todo}(2013)}]{PhysRevE.87.021301}%
  \BibitemOpen
  \bibfield  {author} {\bibinfo {author} {\bibfnamefont {Y.}~\bibnamefont
  {Motoyama}}\ and\ \bibinfo {author} {\bibfnamefont {S.}~\bibnamefont
  {Todo}},\ }\href {\doibase 10.1103/PhysRevE.87.021301} {\bibfield  {journal}
  {\bibinfo  {journal} {Phys. Rev. E}\ }\textbf {\bibinfo {volume} {87}},\
  \bibinfo {pages} {021301(R)} (\bibinfo {year} {2013})}\BibitemShut {NoStop}%
\bibitem [{\citenamefont {Chepiga}\ \emph {et~al.}(2016)\citenamefont
  {Chepiga}, \citenamefont {Affleck},\ and\ \citenamefont
  {Mila}}]{PhysRevB.94.205112}%
  \BibitemOpen
  \bibfield  {author} {\bibinfo {author} {\bibfnamefont {N.}~\bibnamefont
  {Chepiga}}, \bibinfo {author} {\bibfnamefont {I.}~\bibnamefont {Affleck}}, \
  and\ \bibinfo {author} {\bibfnamefont {F.}~\bibnamefont {Mila}},\ }\href
  {\doibase 10.1103/PhysRevB.94.205112} {\bibfield  {journal} {\bibinfo
  {journal} {Phys. Rev. B}\ }\textbf {\bibinfo {volume} {94}},\ \bibinfo
  {pages} {205112} (\bibinfo {year} {2016})}\BibitemShut {NoStop}%
\bibitem [{\citenamefont {Kariyado}\ \emph {et~al.}(2018)\citenamefont
  {Kariyado}, \citenamefont {Morimoto},\ and\ \citenamefont
  {Hatsugai}}]{PhysRevLett.120.247202}%
  \BibitemOpen
  \bibfield  {author} {\bibinfo {author} {\bibfnamefont {T.}~\bibnamefont
  {Kariyado}}, \bibinfo {author} {\bibfnamefont {T.}~\bibnamefont {Morimoto}},
  \ and\ \bibinfo {author} {\bibfnamefont {Y.}~\bibnamefont {Hatsugai}},\
  }\href {\doibase 10.1103/PhysRevLett.120.247202} {\bibfield  {journal}
  {\bibinfo  {journal} {Phys. Rev. Lett.}\ }\textbf {\bibinfo {volume} {120}},\
  \bibinfo {pages} {247202} (\bibinfo {year} {2018})}\BibitemShut {NoStop}%
\bibitem [{\citenamefont {Motoyama}\ and\ \citenamefont
  {Todo}(2018)}]{PhysRevB.98.195127}%
  \BibitemOpen
  \bibfield  {author} {\bibinfo {author} {\bibfnamefont {Y.}~\bibnamefont
  {Motoyama}}\ and\ \bibinfo {author} {\bibfnamefont {S.}~\bibnamefont
  {Todo}},\ }\href {\doibase 10.1103/PhysRevB.98.195127} {\bibfield  {journal}
  {\bibinfo  {journal} {Phys. Rev. B}\ }\textbf {\bibinfo {volume} {98}},\
  \bibinfo {pages} {195127} (\bibinfo {year} {2018})}\BibitemShut {NoStop}%
\bibitem [{\citenamefont {Kawarabayashi}\ \emph {et~al.}(2019)\citenamefont
  {Kawarabayashi}, \citenamefont {Ishii},\ and\ \citenamefont
  {Hatsugai}}]{kawarabayashi2019fractionally}%
  \BibitemOpen
  \bibfield  {author} {\bibinfo {author} {\bibfnamefont {T.}~\bibnamefont
  {Kawarabayashi}}, \bibinfo {author} {\bibfnamefont {K.}~\bibnamefont
  {Ishii}}, \ and\ \bibinfo {author} {\bibfnamefont {Y.}~\bibnamefont
  {Hatsugai}},\ }\href {\doibase /10.7566/JPSJ.88.045001} {\bibfield  {journal}
  {\bibinfo  {journal} {J. Phys. Soc. Jpn.}\ }\textbf {\bibinfo {volume}
  {88}},\ \bibinfo {pages} {045001} (\bibinfo {year} {2019})}\BibitemShut
  {NoStop}%
\bibitem [{\citenamefont {Mizoguchi}\ \emph
  {et~al.}(2019{\natexlab{b}})\citenamefont {Mizoguchi}, \citenamefont
  {Araki},\ and\ \citenamefont {Hatsugai}}]{doi:10.7566/JPSJ.88.104703}%
  \BibitemOpen
  \bibfield  {author} {\bibinfo {author} {\bibfnamefont {T.}~\bibnamefont
  {Mizoguchi}}, \bibinfo {author} {\bibfnamefont {H.}~\bibnamefont {Araki}}, \
  and\ \bibinfo {author} {\bibfnamefont {Y.}~\bibnamefont {Hatsugai}},\ }\href
  {\doibase 10.7566/JPSJ.88.104703} {\bibfield  {journal} {\bibinfo  {journal}
  {Journal of the Physical Society of Japan}\ }\textbf {\bibinfo {volume}
  {88}},\ \bibinfo {pages} {104703} (\bibinfo {year}
  {2019}{\natexlab{b}})}\BibitemShut {NoStop}%
\bibitem [{\citenamefont {Su}\ \emph {et~al.}(1979)\citenamefont {Su},
  \citenamefont {Schrieffer},\ and\ \citenamefont
  {Heeger}}]{PhysRevLett.42.1698}%
  \BibitemOpen
  \bibfield  {author} {\bibinfo {author} {\bibfnamefont {W.~P.}\ \bibnamefont
  {Su}}, \bibinfo {author} {\bibfnamefont {J.~R.}\ \bibnamefont {Schrieffer}},
  \ and\ \bibinfo {author} {\bibfnamefont {A.~J.}\ \bibnamefont {Heeger}},\
  }\href {\doibase 10.1103/PhysRevLett.42.1698} {\bibfield  {journal} {\bibinfo
   {journal} {Phys. Rev. Lett.}\ }\textbf {\bibinfo {volume} {42}},\ \bibinfo
  {pages} {1698} (\bibinfo {year} {1979})}\BibitemShut {NoStop}%
\bibitem [{\citenamefont {Fukui}\ \emph {et~al.}(2005)\citenamefont {Fukui},
  \citenamefont {Hatsugai},\ and\ \citenamefont {Suzuki}}]{fukui2005chern}%
  \BibitemOpen
  \bibfield  {author} {\bibinfo {author} {\bibfnamefont {T.}~\bibnamefont
  {Fukui}}, \bibinfo {author} {\bibfnamefont {Y.}~\bibnamefont {Hatsugai}}, \
  and\ \bibinfo {author} {\bibfnamefont {H.}~\bibnamefont {Suzuki}},\ }\href
  {\doibase 10.1143/JPSJ.74.1674} {\bibfield  {journal} {\bibinfo  {journal}
  {J. Phys. Soc. Jpn.}\ }\textbf {\bibinfo {volume} {74}},\ \bibinfo {pages}
  {1674} (\bibinfo {year} {2005})}\BibitemShut {NoStop}%
\bibitem [{foo({\natexlab{b}})}]{footnote_two}%
  \BibitemOpen
  \href@noop {} {}\bibinfo {note} {Note that, if $\omega_0$ is exactly at the
  frequency of the corner state, the resonance occurs, which results in the
  break down of the approximation of the small displacement}\BibitemShut
  {NoStop}%
\bibitem [{\citenamefont {Yoshino}\ and\ \citenamefont
  {Okazaki}(1977)}]{doi:10.1143/JPSJ.43.415}%
  \BibitemOpen
  \bibfield  {author} {\bibinfo {author} {\bibfnamefont {S.}~\bibnamefont
  {Yoshino}}\ and\ \bibinfo {author} {\bibfnamefont {M.}~\bibnamefont
  {Okazaki}},\ }\href {\doibase 10.1143/JPSJ.43.415} {\bibfield  {journal}
  {\bibinfo  {journal} {J. Phys. Soc. Jpn.}\ }\textbf {\bibinfo {volume}
  {43}},\ \bibinfo {pages} {415} (\bibinfo {year} {1977})}\BibitemShut
  {NoStop}%
\bibitem [{\citenamefont {Janssen}(1998)}]{JANSSEN19981}%
  \BibitemOpen
  \bibfield  {author} {\bibinfo {author} {\bibfnamefont {M.}~\bibnamefont
  {Janssen}},\ }\href {\doibase https://doi.org/10.1016/S0370-1573(97)00050-1}
  {\bibfield  {journal} {\bibinfo  {journal} {Physics Reports}\ }\textbf
  {\bibinfo {volume} {295}},\ \bibinfo {pages} {1 } (\bibinfo {year}
  {1998})}\BibitemShut {NoStop}%
\bibitem [{\citenamefont {Araki}\ \emph {et~al.}(2019)\citenamefont {Araki},
  \citenamefont {Mizoguchi},\ and\ \citenamefont
  {Hatsugai}}]{PhysRevB.99.085406}%
  \BibitemOpen
  \bibfield  {author} {\bibinfo {author} {\bibfnamefont {H.}~\bibnamefont
  {Araki}}, \bibinfo {author} {\bibfnamefont {T.}~\bibnamefont {Mizoguchi}}, \
  and\ \bibinfo {author} {\bibfnamefont {Y.}~\bibnamefont {Hatsugai}},\ }\href
  {\doibase 10.1103/PhysRevB.99.085406} {\bibfield  {journal} {\bibinfo
  {journal} {Phys. Rev. B}\ }\textbf {\bibinfo {volume} {99}},\ \bibinfo
  {pages} {085406} (\bibinfo {year} {2019})}\BibitemShut {NoStop}%
\end{thebibliography}
%\input{main.bbl}

\end{document}